\documentstyle[aps,epsf,twocolumn]{revtex}

\newcommand{\be}{\begin{equation}}

\renewcommand{\vec}[1]{{\bf #1}}
\newcommand{\ee}{\end{equation}}
\newcommand{\bea}{\begin{eqnarray}}
\newcommand{\eea}{\end{eqnarray}}
\newcommand{\vx}{{\bf x}}
\newcommand{\vp}{{\bf p}}

\newcommand{\SKIP}[1]{}
\newcommand{\LC}[1]{\multicolumn{1}{c|}{#1}}
\newcommand{\LCT}[1]{\multicolumn{2}{c|}{#1}}

\begin{document}
\title{Saturation of Elliptic Flow and the Transport Opacity\\
of the Gluon Plasma at RHIC}

\draft
\author{D\'enes Moln\'ar\protect{$^{1,2}$} and Miklos Gyulassy$^{1,2,3}$}

\address{$^1$Physics Department, Columbia University,
         538 W. 120th Street, New York, NY 10027\\
$^2$
RMKI Research Institute for Particle and Nuclear Physics, PO Box 49, H-1525 Budapest, Hungary\\
$^{3}$Collegium Budapest, Szentharomsag u. 2., H-1014 Budapest, Hungary}

\date{Final published version + NPA703 Erratum }

\maketitle

\begin{abstract}
Differential elliptic flow 
and particle spectra are calculated taking into 
account the finite transport opacity of the gluon plasma produced in Au+Au at $E_{cm}\sim 130$ $A$ GeV 
at RHIC. Covariant numerical solutions of the ultrarelativistic 
Boltzmann equation are obtained
using the MPC parton cascade technique.
For  typical pQCD ($\sim 3$ mb) elastic cross sections,
extreme  initial gluon densities, $dN_g/d\eta \sim 15000$,
are required to reproduce the elliptic flow saturation pattern
reported by STAR.
However, we show that the solutions depend mainly
on the transport opacity,
$\chi=\int dz \sigma_t\rho_g$, and thus
the data can also be reproduced with 
$dN_g/d\eta \sim 1000$, but with 
extreme elastic parton cross sections, $\sim 45$ mb.
We demonstrate that the  spectra
and elliptic flow 
are  dominated by numerical artifacts
unless  parton  subdivisions $\sim 100-1000$ are applied
to retain Lorentz covariance  for RHIC initial conditions.
\end{abstract}

\pacs{{\it PACS 12.38.Mh; 24.85.+p; 25.75.-q}}

\section{Introduction}
\label{Section:intro}

Differential elliptic flow, 
$v_2(p_\perp)=\langle \cos(2\phi)\rangle_{p_\perp}$,
 the second Fourier moment of the azimuthal momentum distribution
for fixed $p_\perp$, is one of the important experimental probes of collective
dynamics in $A+A$ reactions
\cite{Stocker:1986ci,Ollitrault:1992bk,Sorge:1997pc,Sorge:1999dm,%
Voloshin:1996mz,Kolbhydro,Csernai:1999nf,Teaney:2000cw,Zhang:1999rs,%
Zabrodin:2000xc,Bleicher:2000sx,Wang:2000fq,Gyulassy:2000gk}.
The  discovery\cite{STARv2}  of a factor of two enhancement
of elliptic flow in noncentral nuclear collisions
at RHIC relative to SPS\cite{Appelshauser:1998dg} 
has generated even more  interest
in this ``barometric'' measure of collective transverse flow.
In addition, preliminary STAR data reported in \cite{Snellings}
suggest a remarkable saturation property of this flow 
at high $p_\perp$ with  $v_2(p_\perp>2 $ GeV)
 $\sim 0.15$.
This corresponds to a factor of two azimuthal angle asymmetry 
of  high-$p_\perp$ particle production relative to the reaction plane.
This collective effect
 depends strongly on the dynamics in a heavy
ion collision and therefore provides  important constraints
about the 
density and effective energy loss of partons.
In particular, we show that it constrains the transport opacity of the produced gluon plasma.

Predictions of collective
elliptic flow in noncentral nuclear collisions
were first based on ideal (nondissipative)
hydrodynamics\cite{Stocker:1986ci,Ollitrault:1992bk,Kolbhydro}.
Unlike at lower energies, ideal hydrodynamics
seems to reproduce the (low $p_T<2$ GeV) data\cite{STARv2}
at RHIC
remarkably well.
However, it fails to saturate at high $p_\perp>2$ GeV 
as indicated by the preliminary data\cite{Snellings}.
The hydrodynamic results were found in \cite{Kolbhydro}
to be surprisingly
insensitive 
to the choice of initial conditions, equation of state and freezeout criteria,
once the observed $dN_{ch}/d\eta$ was reproduced,
leaving no adjustable hydrodynamic model parameters with which
the saturation property could be reproduced.

The lack of saturation in ideal hydrodynamics is due to
the assumption of zero mean free path and
that local equilibrium can be maintained
until a chosen freezeout 3D hypersurface is reached.
This idealization is certainly  invalid outside some
finite domain of phase space 
in heavy ion collisions\cite{nonequil}. 
Finite  transition rates are  expected to produce
nonequilibrium deviations from the predicted hydrodynamic flow pattern.
Covariant  Boltzmann transport theory provides a convenient framework
to estimate dissipative effects. The assumption of local equilibrium
is replaced by the assumption of a finite local mean free path $\lambda(s,x) \equiv 1/\sigma(s) n(x)$. The theory then naturally  
interpolates between  free streaming ($\lambda=\infty$)
and  ideal hydrodynamics ($\lambda=0$).

The previous calculations of collective flow from covariant
transport theory \cite{Zhang:1999rs,molnar:QM99,Molnar:2001yv}
lead to too small collective effects.
This was  due to
the use of  small perturbative QCD cross sections and
 dilute parton initial densities
based on HIJING\cite{Gyulassy:1994ew}. 
Recently, denser  parton initial conditions were suggested
based on  
gluon saturation models\cite{Eskola:2000fc}.
Initial gluon densities up to five times higher
than from HIJING were predicted.
The question studied in this paper
is whether such initial conditions  may  be dense 
enough to generate the observed collective flow
even with pQCD elastic cross sections.
In this paper, 
we explore the dependence of differential elliptic flow on  initial conditions,
or equivalently%
\footnote{
The equivalence is due to the scaling property explained in Section
\ref{Subsection:subdivision}.},
on the magnitude of partonic cross section.

We note that most hadronic cascade models supplemented with string
dynamics\cite{Sorge:1999dm,Bleicher:2000sx}
underpredict elliptic flow
because the spatial asymmetry is too small
after hadronization to generate the observed momentum space asymmetry.
The reason is that
hadronization through longitudinal string excitations reduces the strength of 
partonic level elliptic flow.
To produce sufficient elliptic flow in string models,
other mechanisms have to be included
such as color exchange.
While the impact parameter dependence of
elliptic flow at RHIC can be reproduced via color exchange
\cite{Zabrodin:2000xc},
it is not known whether the saturation property studied here
can also be explained.

The saturation and eventual decrease 
of $v_2$ at high $p_\perp$ has been predicted 
as a consequence of finite inelastic parton 
energy loss\cite{Wang:2000fq,Gyulassy:2000gk}. Those predictions
however assumed the validity of an Eikonal
approach  at moderate $p_T\sim 10$ GeV. In addition, 
the pQCD computable jet quenched part had  to be  
joined phenomenologically onto a parametrized ``soft'' 
nonperturbative component below $p_T<2$ GeV/c.
Covariant transport theory overcomes the need to treat soft and hard dynamics
on different footings. It is the only practical self-consistent theoretical
tool at present 
to address simultaneously both the soft collective component
and the far from equilibrium high-$p_\perp$ component.
While current parton cascade techniques
lack at present a  {\em practical}  means to implement covariant
inelastic energy loss, it is of considerable theoretical interest to solve
covariant Boltzmann theory even in the elastic limit since so few
solutions are known. We solve that theory numerically here 
 to get insight into the
dynamical interplay between the soft and hard  components over a wide
dynamical range of parameters.

For large enough elastic opacities, the observed collective flow strength
can certainly be reproduced \cite{Molnar:2001yv}.
The outstanding question which we focus on is
whether the detailed pattern of deviations from
ideal  hydrodynamic flow and its saturation
at high $p_\perp$ can also be understood quantitatively
in this particular dynamical framework.

Forerunners of this study
\cite{Zhang:1999rs,Molnar:2001yv}
computed elliptic flow
for partonic systems starting from initial conditions expected at RHIC.
In this paper we extend those studies in three aspects.
We compute the $p_\perp$-differential elliptic flow  $v_2(p_\perp)$.
The consequences of  hadronization  are investigated,
which is necessary to compare to  the observations.
Finally, we use realistic diffuse nuclear geometry for the initial conditions.

We compute the partonic evolution with MPC\cite{MPC},
a newly formulated, covariant, parton kinetic theory technique.
MPC is 
an extension of the covariant parton cascade algorithm,
ZPC\cite{Zhang:1998ej}.
Both MPC and ZPC have been
extensively tested\cite{Gyulassy:1997ib,Zhang:1998tj} 
and compared to analytic transport solutions
and covariant Euler and Navier-Stokes dynamics in 1+1D geometry.
A critical feature  of both these algorithms
is the implementation  of the
parton subdivision technique proposed by Pang\cite{Zhang:1998tj,Yang}.

Extensions of MPC 
to include inelastic $2\leftrightarrow 3$ partonic
processes\cite{molnar:QM99} are under development.
In this paper, we apply  MPC
in the pure elastic parton interactions mode as in ZPC\cite{OSCAR}.
 
\section{Covariant Transport Theory}
\label{Section:transport_theory}

\subsection{Transport equation}
\label{Subsection:transport_eqn}

We consider here, as in Refs. \cite{Yang,Zhang:1998ej,MPC,nonequil},
the simplest but nonlinear
form of Lorentz-covariant Boltzmann transport theory
in which the on-shell phase space density $f(x,\vp)$,
evolves with an elastic $2\to 2$ rate as
\bea
p_1^\mu \partial_\mu f_1 &=&\int\limits_2\!\!\!\!
\int\limits_3\!\!\!\!
\int\limits_4\!\!
\left(
f_3 f_4 - f_1 f_2
\right)
W_{12\to 34} \delta^4(p_1+p_2-p_3-p_4)
\nonumber \\
&&
+  \, S(x, \vp_1) .
\label{Eq:Boltzmann_22}
\eea
Here $W$ is the square of the scattering matrix element,
the integrals are shorthands
for $\int\limits_i \equiv \int \frac{g\ d^3 p_i}{(2\pi)^3 E_i}$,
where $g$ is the number of internal degrees of freedom,
while $f_j \equiv f(x, \vp_j)$.
The initial conditions are specified by the source function $S(x,\vp)$,
which we discuss at the end of this Subsection.

For our applications below,
we  interpret  $f(x,\vp)$ as describing
an ultrarelativistic massless gluon gas 
with $g=16$ (8 colors, 2 helicities).
We neglect quark degrees of freedom because at RHIC gluons are more abundant.

In principle,
the transport equation (\ref{Eq:Boltzmann_22}) could be extended for bosons
with the substitution $f_1 f_2 \to f_1 f_2 (1+f_3) (1+f_4)$
and a similar one for $f_3 f_4$.
In practice, no covariant algorithm yet exists
to handle such nonlinearities.
We therefore limit our study to quadratic dependence of the collision
integral on $f$.

The elastic gluon scattering matrix elements in dense parton systems
are modeled by a Debye-screened form\ as in Ref. \cite{Zhang:1999rs}:
\be
\frac{d\sigma_{el}}{dt}
  = \sigma_0 (s)
    \left(1 + \frac{\mu^2}{s}\right)
    \frac{\mu^2}{\left(t+\mu^2\right)^2},
\label{Eq:cross_section}
\ee
where $\mu$ is the screening mass,
$\sigma_0(s) = 9\pi\alpha_s^2(s)/2\mu^2$
is the total cross section.
For simplicity, we will assume $\sigma_0$ to be energy independent,
 neglecting its weak logarithmic dependence on $s$ in the energy range
relevant at RHIC.

For small values of $\mu$, forward-peaked scattering is favored,
while as $\mu$ increases the angular distribution
becomes more and more isotropic.
For a fixed total cross section%
\footnote{
To keep the total cross section constant as a function of $\mu$,
one of course has to adjust the coupling $\alpha_s$ accordingly.},
the relevant transport cross section is
\bea
\sigma_t(s) &\equiv& \int d\sigma_{el} \sin^2\theta_{cm}=\int
dt\, \frac{d\sigma_{el}}{dt}\frac{4t}{s}\left(1-\frac{t}{s}\right)
\nonumber \\
&=&\sigma_0 4z(1+z)\left[(2z+1)\ln(1+1/z) - 2\right]
\;\;,
\label{Eq:transport_xs}
\eea
where $z\equiv\mu^2/s$.
This is a monotonic function of $\mu$
and maximal in the isotropic ($\mu\to\infty$) case.
In the small angle dominated limit, with $z\ll 1$,
$\sigma_t/\sigma_0 \approx  4z(\ln 1/z - 2)$.

It is important to emphasize that while the cross section suggests a
geometrical picture of action over finite distances, we use
Eq. (\ref{Eq:cross_section}) only as a convenient parametrization to
describe the effective {\em local} transition probability, $W$.  In
the present study this is simply modeled as
$W(t)=s\;d\sigma/dt$.
The particle subdivision technique
(see next Subsection) needed to recover covariance removes all notion of
nonlocality in this approach, just like in hydrodynamics.  Thus, the
cross sections, e.g., 100 mb, used in the present study to simulate
a high collision rate do {\em not} imply acausal action at a distance.

For Au+Au collisions,
the initial condition was taken to be
a longitudinally boost invariant Bjorken tube
in local thermal equilibrium
at temperature $T_0$ at proper time $\tau_0=0.1$~fm/$c$
with uniform  pseudorapidity $\eta\equiv 1/2 \log((t+z)/(t-z))$ distribution
between $|\eta| < 5$.
The transverse density distribution was
assumed to be proportional
to the binary collision
distribution for two Woods-Saxon distributions.
For collisions at impact parameter $b$
the  transverse binary collision profile is  
\be
\frac{dN({\bf b})}{d\eta d^2 {\bf x}_\perp}= \sigma_{jet}
T_A\!\!\left({\bf x}_\perp +\frac{{\bf b}}{2}\right)
T_A\!\!\left({\bf x}_\perp -\frac{{\bf b}}{2}\right)
\; \; ,
\label{Eq:tab}
\ee
where $T_A({\bf b})\equiv \int dz \rho_A(\sqrt{z^2+{\bf b}^2})$, in terms of the diffuse nuclear density $\rho_A(r)$.
The pQCD jet cross section
normalization,
$\sigma_{jet}$, and the temperature $T_0$
were determined
by fitting the gluon minijet transverse momentum spectrum
predicted by HIJING\cite{Gyulassy:1994ew}
for central Au+Au collisions at $\sqrt{s} = 130A$ GeV
(without shadowing and jet quenching).
This gives $dN(0)/d\eta = 210$ and $T_0 = 700$ MeV.

Evolutions from different initial densities (but the same density profile)
can be obtained by varying the cross section only
and using the scaling property explained in Section \ref{Subsection:scaling}.

\subsection{Parton Subdivision}
\label{Subsection:subdivision}

We utilize the parton cascade method
to solve the Boltzmann transport equation (\ref{Eq:Boltzmann_22}).
A critical drawback of all cascade algorithms is
that they inevitably lead to numerical artifacts
that violate Lorentz covariance.
This occurs because particle interactions are simulated to occur
whenever the distance of closest approach (in the relative c.m.)
is $d<\sqrt{\sigma_0/\pi}$.

Acausal (superluminal)
propagation due to
action at a distance leads to severe numerical artifacts.
In particular,
the transverse energy evolution $dE_T(\tau)/dy$ and the final asymptotic
transverse energy per unit rapidity are frame dependent\cite{nonequil}.

To recover the {\em local} character of equation (\ref{Eq:Boltzmann_22})
and hence Lorentz covariance,
it is essential to use
the parton subdivision technique\cite{Yang,Zhang:1998ej}.
This technique is based  on the covariance of Eq. (\ref{Eq:Boltzmann_22})
under the transformation 
\be f\to f'\equiv \ell\, f, \quad W\to W'\equiv W/\ell \quad
(\sigma\to \sigma' = \sigma/ \ell) \ ,
\label{Eq:particle_subdivision}
\ee
where $\ell$ is the number of particle subdivisions.
The magnitude of numerical artifacts
is governed by the diluteness of the system
$\sqrt{\sigma}/\lambda_{MFP}$,
which scales with $1/\sqrt{\ell}$ \cite{Zhang:1998tj}.
Lorentz violation therefore formally vanishes in the $\ell\to \infty$ limit.
The convergence to the accurate covariant solution with $\ell$ 
is {\em slower} if the density or cross section increases.

\begin{figure}[hp]
\center
\leavevmode
    \epsfysize 6cm
    \epsfbox{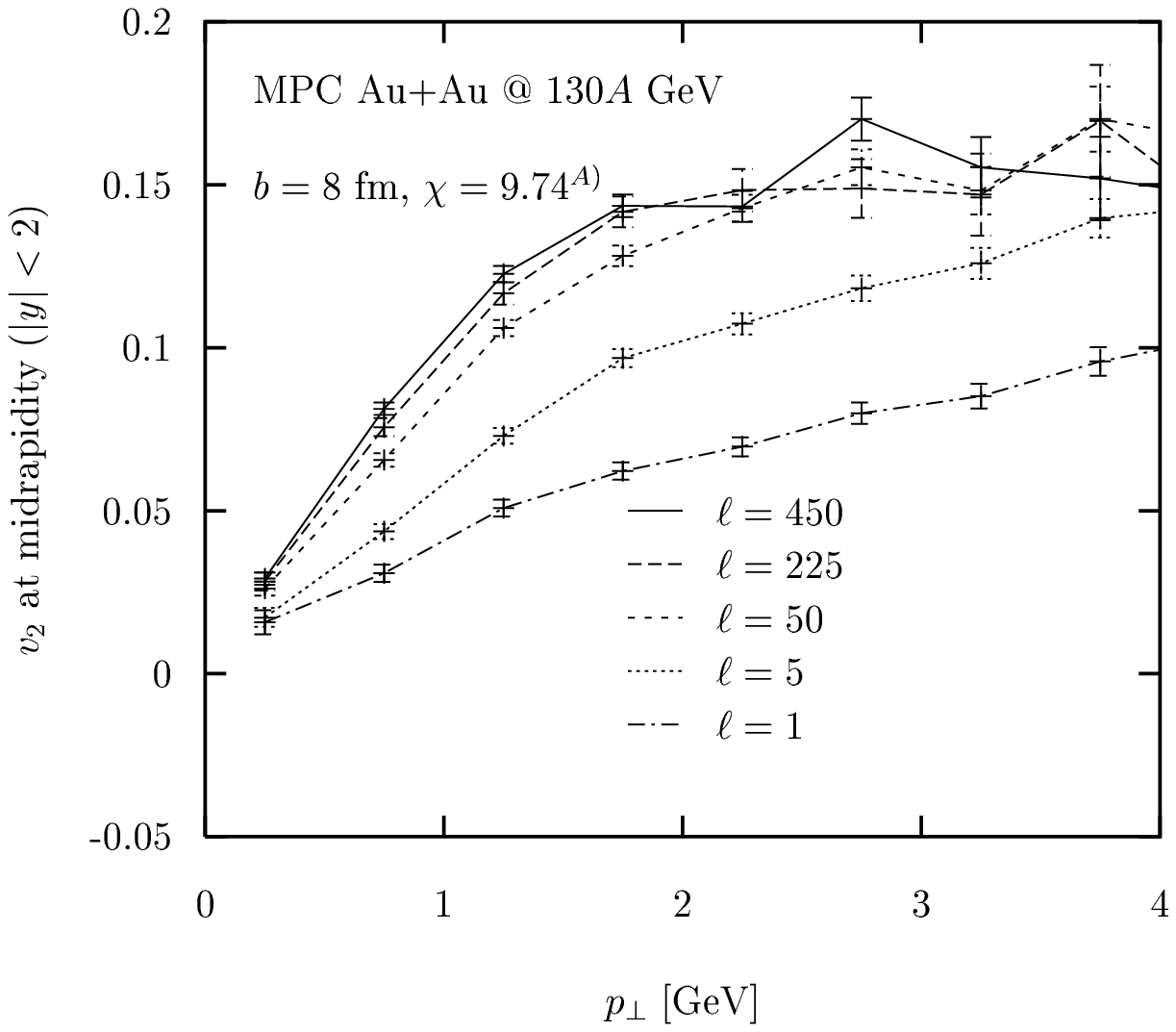}
\caption{
\footnotesize
Strong dependence of the gluon
elliptic flow\ on parton subdivision
as a function of $p_\perp$\ is shown 
for Au+Au at $\sqrt{s}=130A$ GeV with $b=8$ fm. 
Solutions for transport opacity $\chi = 9.74^{A)}$ (see Table \ref{Table:chi}),
and particle subdivisions
$\ell=1$, 5, 50, 225, and 450\ are shown.
}
\label{Figure:1}
\label{Figure:v2_vs_l}
\end{figure}

\begin{figure}[hp]
\center
\leavevmode
\hbox{
    \epsfysize 8cm
    \epsfbox{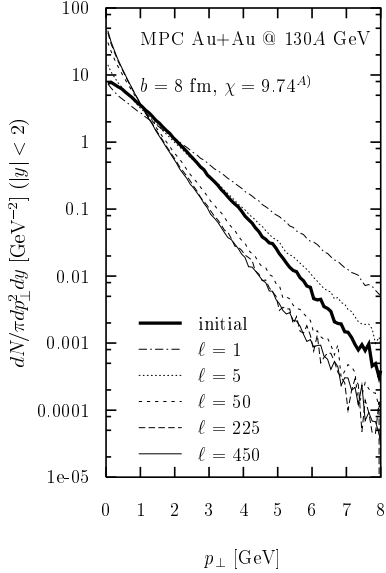}
}
\caption{
\footnotesize
Strong dependence of the 
final gluon $p_\perp$ spectra\ on parton subdivision
is shown for Au+Au at $\sqrt{s}=130A$ GeV with $b=8$ fm.
Solutions for transport opacity $\chi = 9.74^{A)}$
(see Table \ref{Table:chi}),
and
particle subdivisions
$\ell=1$, 5, 50, 225, and 450\ are shown as in Fig.
\protect\ref{Figure:v2_vs_l}.
The spectra are normalized here to $dN(0)/d\eta = 210$.
}
\label{Figure:2}
\label{Figure:pt_vs_l}
\end{figure}

Figures \ref{Figure:v2_vs_l} and \ref{Figure:pt_vs_l} illustrate
the effect of Lorentz violation on the observables
studied in this paper.
For insufficient particle subdivision,
elliptic flow and the $p_\perp$ spectra are dominated by numerical artifacts.
In particular,
elliptic flow is significantly underpredicted,
while the high-$p_\perp$ spectrum exhibits an unphysical ``reheating''
during the expansion.
For the initial conditions for these plots,
these numerical artifacts disappear 
only when the particle subdivision reaches $\sim 200$.
This reinforces the results by Ref. \cite{nonequil},
where very high $\sim 100-1000$ subdivisions
were needed to obtain accurate numerical solutions
of the transport equation (\ref{Eq:Boltzmann_22})
for initial conditions expected at RHIC.

\subsection{Scaling of the transport solutions}
\label{Subsection:scaling}

Subdivision covariance (\ref{Eq:particle_subdivision})
actually implies that the transport equation has a broad dynamical range,
and the solution for any given initial condition and transport property
immediately provides the solution to a broad band
of suitably scaled initial conditions and transport properties. 
This is because solutions
for problems with $\ell$ times larger the initial density
$dN/d\eta d^2x_\perp$,
but with one $\ell$-th the reaction rate
can be mapped to the original ($\ell=1$) case for {\em any} $\ell$. 
We must use subdivision to eliminate numerical artifacts.
However, once that is achieved,
we have actually found the solution to a whole class of 
suitably rescaled problems. 

The dynamical range of the transport equation (\ref{Eq:Boltzmann_22}) 
is further increased by its covariance under coordinate
and momentum rescaling\cite{nonequil},
leading to covariance of the transport theory under 
\bea
f(x,\vec p) &\to& f'(x,\vec p)
\equiv \ell_p^{-3} \ell \,f\!\left(\frac{x}{\ell_x},\frac{\vec p}{\ell_p}\right),
\nonumber\\
W(\{p_i\}) &\to& W'(\{p_i\})
\equiv \frac{\ell_p^2}{\ell_x \ell}\,
W\!\left(\left\{\frac{p_i}{\ell_p}\right\}\right),
\nonumber\\
m&\to& m'=m/\ell_p,
\label{Eq:rescale_all}
\eea
where $\ell_x$ and $\ell_p$ are the coordinate and momentum scaling parameters,
respectively.
This means \cite{nonequil}
that we can scale one solution to others 
provided that both
$\mu/T_0$ and $\sigma_0 dN/d\eta$ remain the same
(we cannot exploit coordinate
scaling because the nuclear geometry is fixed).
For example,
three times the density with one-third the cross section
leaves both parameters the same,
hence the results can be obtained via scaling
without further computation.

In general  the numerical 
(cascade) solution of Eq. (\ref{Eq:Boltzmann_22}) tends 
in the $\ell\rightarrow \infty$ limit toward a covariant physical
solution that depends on two scales,  
$\mu/T_0$ and $\sigma_0 dN/d\eta$. In an Eikonal picture of high-$p_\perp$
production, the distributions are expected to depend on
 the opacity or the mean number of collisions in the medium 
\bea
\langle n\rangle &=&\frac{L}{\lambda_{el}}= \int
dt\, \frac{d\sigma_{el}}{dt}\int dz 
\rho\left(\vx_0+z \hat{\bf n},\tau=\frac{z}{c} \right)\nonumber \\
 &\approx & \, \frac{dN}{dy} \,
\frac{\sigma_0}{2 \pi R_G^2} \, \log \frac{R_G}{\tau_0} \;,
\eea
where $\tau_0$ is the formation proper time 
and $R_G$ is the effective Gaussian transverse coordinate rms radius.

Our numerical results in Table \ref{Table:chi} show
that for a given centrality and initial gluon density,
the average number of collisions per parton is
within 10\% accuracy
proportional to $\sigma_0$ and does not depend on
$\mu/T_0$:
\be
\langle n(b,\sigma_0 
\frac{dN(0)}{d\eta},\frac{\mu}{T_0}) \rangle
\approx \frac{\sigma_0}{\sigma_0'}  
\;\langle n(b,\sigma_0' 
\frac{dN(0)}{d\eta},\frac{\mu'}{T_0'}) \rangle \ .
\label{Eq:compute_n}
\ee
Therefore, one would naively expect $\sigma_0 dN/d\eta$ to be
the relevant scale.

However, from the point of view of dissipative dynamics
via Navier-Stokes and Fokker-Planck
equation, the more relevant dynamical 
 parameter is the
effective {\em transport opacity} 
\be
\chi\equiv \frac{\sigma_{t}}{\sigma_0} \langle n \rangle
= \sigma_t
\langle \int dz  
\rho\left({\bf x}_0+ z\hat{\bf n},\tau=\frac{z}{c} \right)\rangle
\;\; .\label{Eq:tropacity}
\ee
The ensemble average over initial coordinates and directions is implied above.
In addition,
$\sigma_t$ here stands for $\sigma_t \equiv \sigma_t(\langle s \rangle)$,
where $\langle s \rangle=18 T_0^2$ is the  initial thermal average of $s$. 
This is an approximation to the ensemble average $\langle \sigma_t(s)\rangle$.

In general, the transport opacity is a {\rm dynamical} quantity
that we do not know until we solved the transport
equation for the set of parameters  $b$, $\sigma_0 dN(0)/d\eta$,
and $\sigma_t / \sigma_0$ (or equivalently, $\mu / T_0$).
However,
Eqs. (\ref{Eq:compute_n})
and (\ref{Eq:tropacity}) imply that
for the range of parameters considered in this study
it is approximately proportional
to the product of the two scales
$\sigma_0 dN(0)/d\eta$ and $\sigma_t/\sigma_0$:
\be
\chi(b,\sigma_0 \frac{dN(0)}{d\eta},\frac{\sigma_t}{\sigma_0}) \approx
\frac{\sigma_t \frac{dN(0)}{d\eta}}{\sigma_0' \frac{dN'(0)}{d\eta}}
 \; 
\langle n(b,\sigma_0' \frac{dN'(0)}{d\eta})\rangle  \ .
\label{Eq:compute_chi}
\ee

The nontrivial, impact parameter dependent part 
 $\langle n(b, \sigma_0 dN(0)/d\eta \rangle$
is the average
number of collisions per parton as a function of $b$
for a fixed value of $\sigma_0 dN(0)/d\eta$
(and an arbitrary $\mu/T_0$),
which is tabulated in Table \ref{Table:chi}.
For example, one could use the
 values of set  E),
in which case the corresponding proportionality constant is 
$\sigma_0' dN'(0)/d\eta = 3{\rm \ mb} \times 210 = 630 {\rm \ mb}$.

\begin{figure}[hp]
\center
\leavevmode
\vbox{
    \epsfysize 6cm
    \epsfbox{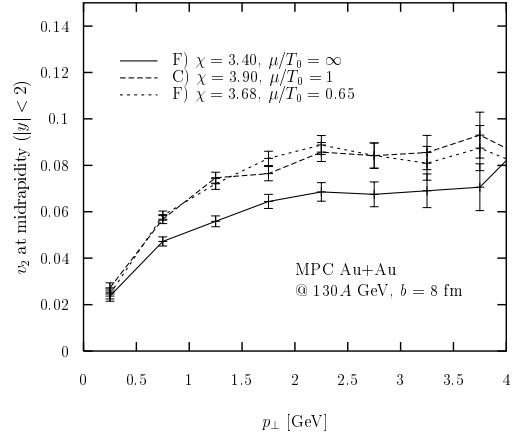}
    \epsfysize 6cm
    \epsfbox{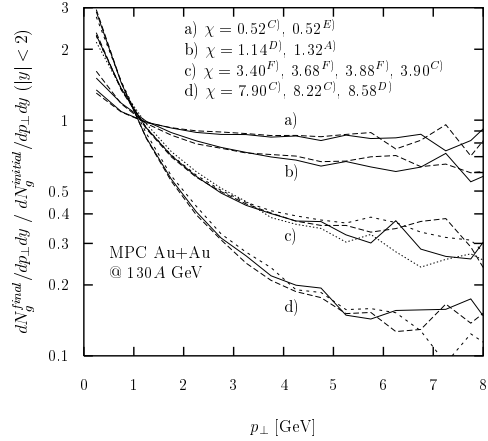}
}
\caption{
\footnotesize
The very weak dependence of the elliptic flow and gluon spectra
quenching on the angular distribution of the parton cross section
is shown.
See Table \ref{Table:chi} for the simulation parameters
corresponding to each curve. The solutions are seen to depend mainly on the
transport opacity.
}
\label{Figure:3}
\label{Figure:verify_transport}
\end{figure}

Of course, there is no a priori guarantee
that the solutions of Eq. (\ref{Eq:Boltzmann_22}) depend
only on this transport opacity parameter.
However,
as demonstrated in Fig. \ref{Figure:verify_transport},
this turns out to hold  within $\sim 10-20 \%$ accuracy
for elliptic flow and the transverse momentum spectra 
 out to 6 GeV/c for the parameters and initial conditions appropriate
at RHIC energies.
In particular,
simulations with very different impact parameters give the same ratio
between the initial and final spectrum, provided $\chi$ is the same.
Hence,
the relevant parameter that governs the evolution of the $p_\perp$ spectra
is $\chi$ alone.
This is not so for elliptic flow,
which is also driven by the initial spatial anisotropy
and thus depends not only on $\chi$ but also on the impact parameter.

\section{Numerical results for the partonic evolution}
\label{Section:glue_results}

In this Section we present elliptic flow results and $p_\perp$ spectra for
the partonic evolution.

Under the scaling (\ref{Eq:rescale_all})
the differential elliptic flow $v_2(p_\perp)$ and the $p_\perp$ spectrum
transform as
\bea
v_2(p_\perp) &\to& v_2'(p_\perp)
\equiv v_2\!\!\left(\frac{p_\perp}{\ell_p}\right)
\nonumber\\
\frac{dN}{d^2 p_\perp} (\vec p_\perp)
 &\to& \frac{dN}{d^2 p_\perp}'(\vec p_\perp)
     = \ell \frac{dN}{d^2 p_\perp}\!\!\left(\frac{\vec p_\perp}{\ell_p}\right).
\label{Eq:v2_pt_scaling}
\eea
Hence elliptic flow depends on $\sigma_t$ and $dN/d\eta$
only through the product $\sigma_t dN/d\eta$.
On the other hand, the $p_\perp$ spectrum depends on
$\sigma_t$ and $dN/d\eta$ {\em separately}.

As emphasized in the previous Section,
we will label the results
by the effective elastic transport opacity $\chi$
from Eq. (\ref{Eq:tropacity}) and the impact parameter $b$.
Possible initial gluon densities and transport cross sections
corresponding to a given transport opacity $\chi$ and impact parameter $b$
can be extracted using Eq. (\ref{Eq:compute_chi}),
while possible cutoffs and total cross sections corresponding to a given
transport cross section can be obtained from Eq. (\ref{Eq:transport_xs}).
These mappings are not unique.
A given $\chi$ and $b$ correspond to a whole class of possible
initial densities, total cross sections and cutoffs.

Table \ref{Table:chi}
shows the set of simulation parameters for each simulation,
together with $\chi$
determined directly from
the average number of cascade collisions per particle.
In Table \ref{Table:chi}
we also introduced letter codes A) through F) as a quick reference
to particular subsets of simulation parameters.
We will include this letter code on most labels together with $\chi$,
for convenience.

The evolution was performed numerically 
with 40 and 100 mb isotropic cross sections,
and with 3, 40 and 100 mb gluonic cross sections with $\mu/T_0 = 1$.
We used particle subdivision
$\ell=100$ for impact parameters 0, 2, and 4 fm,
while $\ell=220$, 450, 1100, and 5000,
for $b=6$, 8, 10, and 12 fm.
Our study of subdivision convergence
shown in Figs. \ref{Figure:v2_vs_l} and  \ref{Figure:pt_vs_l}
indicates that with $b=8$ fm and $\chi = 9.74^{A)}$
a particle subdivision of $\ell \sim 200-250$ is required,
which means that for $b=0$, 2, and 4 fm, $\ell=100$ is not sufficient
if $\chi > 8-10$.
Unfortunately, our computational resources were insufficient
to allow  higher subdivision runs.
While this affects elliptic flow results little
because most elliptic flow contributions come from $b > 4$ fm,
the particle spectra are affected significantly.
Therefore, we only present spectra for $b \ge 6$ fm.

\subsection{Elliptic flow results}
\label{Subsection:glue_flow}
\begin{figure}[hp]
\center
\leavevmode
    \epsfysize 6cm
    \epsfbox{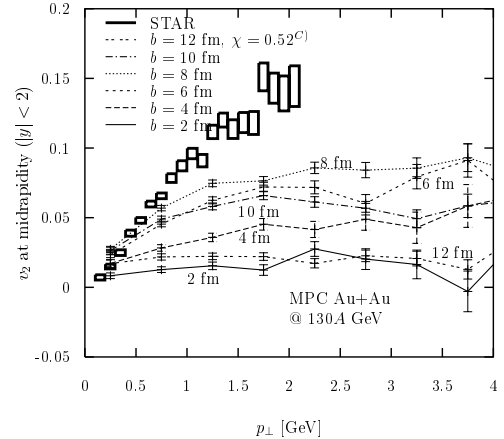}
\caption{
\footnotesize
Gluon elliptic flow as a function of $p_\perp$ 
for Au+Au at $\sqrt{s}=130A$ GeV
with impact parameters $b=2$, 4, 6, 8, 10, and 12 fm\ is shown
for transport opacities $\chi=7.90^{C)}$, 6.98$^{C)}$,
5.68$^{C)}$, 3.90$^{C)}$,
1.98$^{C)}$, and 0.52$^{C)}$.
STAR 
data\protect\cite{STARv2} below 2 GeV/$c$ are shown.
Preliminary STAR data\protect\cite{Snellings} suggest that
$v_2\sim 0.15-0.17$ may saturate in the $2<p_T<4$ GeV/$c$ range.
}
\label{Figure:4}
\label{Figure:v2_g40mb}
\end{figure}

\begin{figure}[hp]
\center
\leavevmode
    \epsfysize 6cm
    \epsfbox{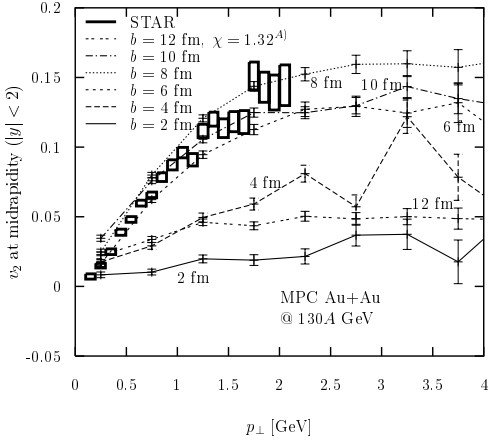}
\caption{
\footnotesize
Same as Fig. \protect\ref{Figure:v2_g40mb} except that solutions for
transport opacities $\chi=19.4^{A)}$, 17.2$^{A)}$,
14.1$^{A)}$, 9.74$^{A)}$,
5.00$^{A)}$, and 1.32$^{A)}$ are shown.
STAR 
data \protect\cite{STARv2} below 2 GeV/$c$ are shown.
Preliminary STAR data\protect\cite{Snellings} suggest that
$v_2\sim 0.15-0.17$ may saturate in the $2<p_T<4$ GeV/$c$ range.
}
\label{Figure:5}
\label{Figure:v2_g100mb}
\end{figure}

\begin{figure}[hp]
\center
\leavevmode
    \epsfysize 6cm
    \epsfbox{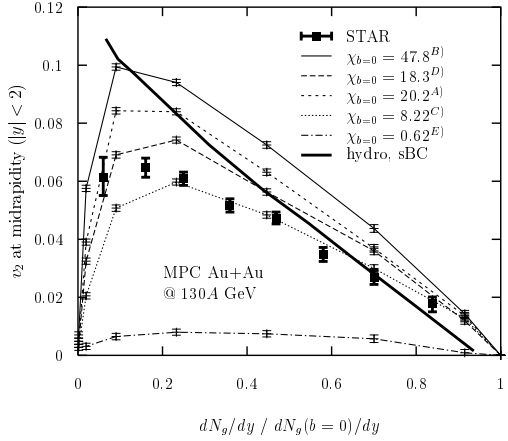}
\caption{
\footnotesize
Gluon elliptic flow as a function of centrality
for Au+Au at $\sqrt{s}=130A$ GeV is shown for
transport opacities $\chi_{b=0} = 0.62^{E)}$, 8.22$^{C)}$,
18.3$^{D)}$,
20.2$^{A)}$, and 47.8$^{B)}$
for $b=0$.
Identical to the charged hadron elliptic flow
as a function of $n_{ch}/n_{ch}^{max}$,
if the gluons are hadronized via local parton-hadron duality.
The ideal hydrodynamics result is taken
from {\protect\cite{Kolbhydro}} with the so called sBC initial conditions.
STAR data\protect\cite{STARv2} are also shown.
}
\label{Figure:6}
\label{Figure:v2_vs_centrality}
\end{figure}

Figures \ref{Figure:v2_g40mb} and \ref{Figure:v2_g100mb}
show the final asymptotic gluon elliptic flow as a function
of transverse momentum for different impact parameters
with $\sigma_t dN(0)/d\eta \approx 2580^{C)}$ mb  and 6440$^{A)}$ mb,
respectively (the corresponding transport opacities at $b=0$ are
$\chi_{b=0} = 7.90^{C)}$ and $19.4^{A)}$).
With increasing $p_\perp$,
elliptic flow increases until $p_\perp\sim 1.5-2$ GeV,
where it saturates,
reproducing the pattern observed at RHIC\cite{STARv2,Snellings}.
With increasing impact parameter,
elliptic flow first monotonically increases,
then monotonically decreases,
showing a maximum at $b\approx 8$ fm.
These features were universal for all the cross sections we studied, 
except for a small increase in the location of the maximum
with increasing transport cross section,
as can be seen in Fig. \ref{Figure:v2_vs_centrality},
from $b = 7$ fm at $\sigma_t = 0.91^{E)}$ mb to 9 fm at $66^{B)}$ mb.
Also,
as expected,
elliptic flow is a monotonically increasing function of the transport opacity,
if the impact parameter is kept fixed.

Figure \ref{Figure:v2_vs_centrality} shows the $p_\perp$-integrated
gluon elliptic flow as a function of centrality.
The cascade reproduces the trend seen in the STAR data\cite{STARv2}
down to  
very small centralities $\sim 0.1-0.2$,
where the ideal hydrodynamical assumption of zero mean free path
certainly breaks down.
To {\em quantitatively} reproduce the data, transport opacities
$\chi_{b=0} \sim 8-14$ are needed.
With the pQCD elastic $gg$ cross section $\sigma_0(\mu=T) \approx 3$ mb,
this corresponds to an initial gluon density $dN_g(0)/d\eta \sim 3000-5000$.

Figure \ref{Figure:minbias_v2} shows the impact-parameter-averaged {\em gluon}
elliptic flow as
a function of transverse momentum for different transport opacities.
The impact-parameter-averaged flow was computed via the formula
\be
v_2^{av}(p_\perp) \equiv \frac{2\pi}{\pi b_{max}^2} \int_0^{b_{max}} db\; b\;
\frac{\int d\phi \cos(2\phi) \frac{dN}{d\eta d^2p_\perp}(b) }
     {\int d\phi \frac{dN}{d\eta d^2p_\perp}(b)}\ ,
\label{Eq:v2MPC}
\ee
with $b_{max} = 12$ fm.
As we show in the Appendix, for our transport theory solutions, 
Eq. (\ref{Eq:v2MPC})
 gives  comparable results to the minimum-bias differential elliptic flow
defined by STAR as
\be
v_{2}^{STAR}(p_\perp)
 \equiv \frac{\int_0^{b_{max}} b\, db
              \int d\phi \cos(2\phi) \frac{dN}{d\eta d^2p_\perp}(b)
             }
             {\int_0^{b_{max}} b\, db
              \int d\phi \frac{dN}{d\eta d^2p_\perp}(b)
             },
\label{Eq:v2STAR}
\ee
which weights flow in more central events preferentially.
This is not the case for ideal hydrodynamic solutions\cite{Kolbhydro},
for which the STAR definition\cite{STARv2}
 results in much smaller flow than if 
Eq. (\ref{Eq:v2MPC}) is used.

We use the definition (\ref{Eq:v2MPC}) because 
it can be computed numerically more reliably in our approach.
As discussed in the beginning of Section \ref{Section:glue_results},
it is difficult to reduce
cascade numerical artifacts to an acceptable level
when the transport opacity $\chi$ is large.
For all other parameters kept fixed,
$\chi$ increases with decreasing impact parameter,
hence the STAR definition that weights $v_2$ at small $b$ preferentially
is more prone to such numerical artifacts.

Varying the magnitude of energy loss
we searched 
for the drop in $v_2(p_\perp)$ at high $p_\perp$
predicted by calculations based on inelastic parton energy loss
\cite{Wang:2000fq,Gyulassy:2000gk}.
Although those studies consider only effects due to radiative energy loss,
one expects a similar behavior in case of purely elastic energy loss.

\begin{figure}[hp]
\center
\leavevmode
    \epsfysize 6cm
    \epsfbox{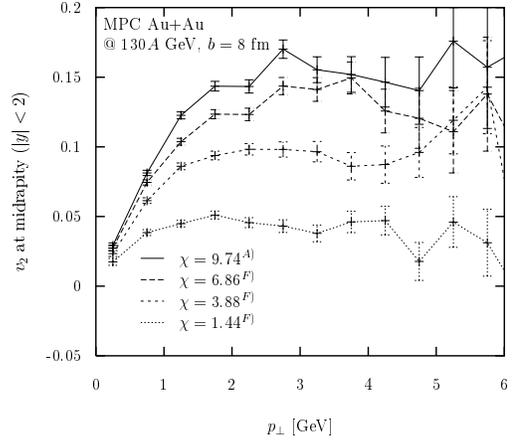}
\caption{
\footnotesize
Gluon elliptic flow as a function of $p_\perp$\ is shown
for Au+Au at $\sqrt{s}=130A$ GeV with $b=8$ fm and 
$\mu/T_0= 0.226$, 0.45, 0.71, and 1
(transport opacities $\chi=1.44^{F)}$, 3.88$^{F)}$,
6.86$^{F)}$, and 9.74$^{A)}$).
}
\label{Figure:8}
\label{Figure:v2_vs_mu}
\end{figure}

Figure \ref{Figure:v2_vs_mu} shows the dependence of $v_2(p_\perp)$ on
the transport opacity for a fixed impact parameter.
We varied the opacity by changing the screening mass $\mu$.
As expected, elliptic
flow decreases with decreasing $\chi$. 
However, there is no sign of a drop at high $p_\perp$:
within statistical errors,
the results are consistent with a constant flow from 2 to 6 GeV
transverse momentum.

\subsection{Particle spectra}
\label{Subsection:glue_spectra}

\begin{figure}[hp]
\center
\leavevmode
\hbox{
    \epsfysize 6cm
    \epsfbox{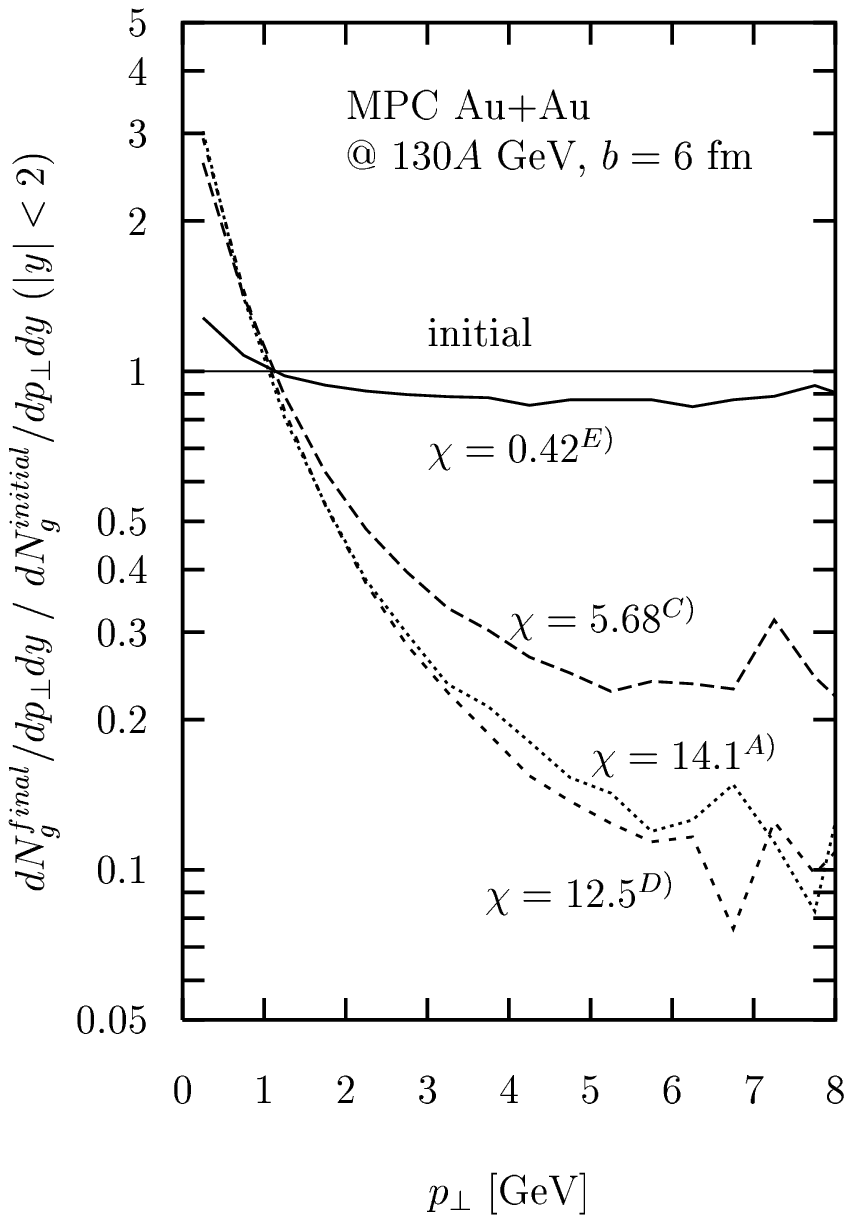}
    \epsfysize 6cm
    \epsfbox{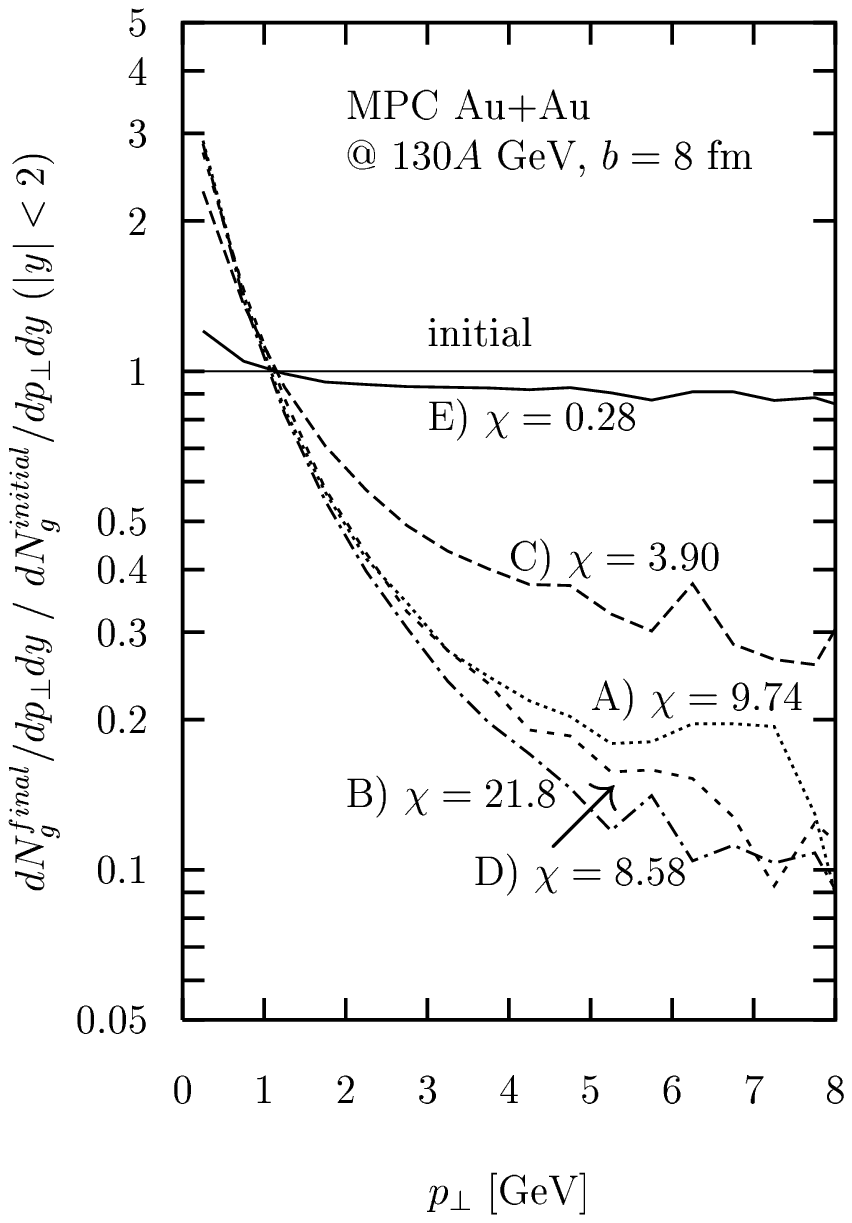}
}
\caption{
\footnotesize
Final gluon $p_\perp$ spectra relative to the thermal initial spectrum
are shown
as a function of transport opacity
for Au+Au at $\sqrt{s} = 130A$ GeV
with $b=6$ fm (left) and $b=8$ fm (right).
Proportional to the charged hadron $p_\perp$ spectra
if the gluons are hadronized via local parton-hadron duality.
}
\label{Figure:9}
\label{Figure:pt_vs_sigma-1}
\end{figure}

\begin{figure}[hp]
\center
\leavevmode
\hbox{
    \epsfysize 6cm
    \epsfbox{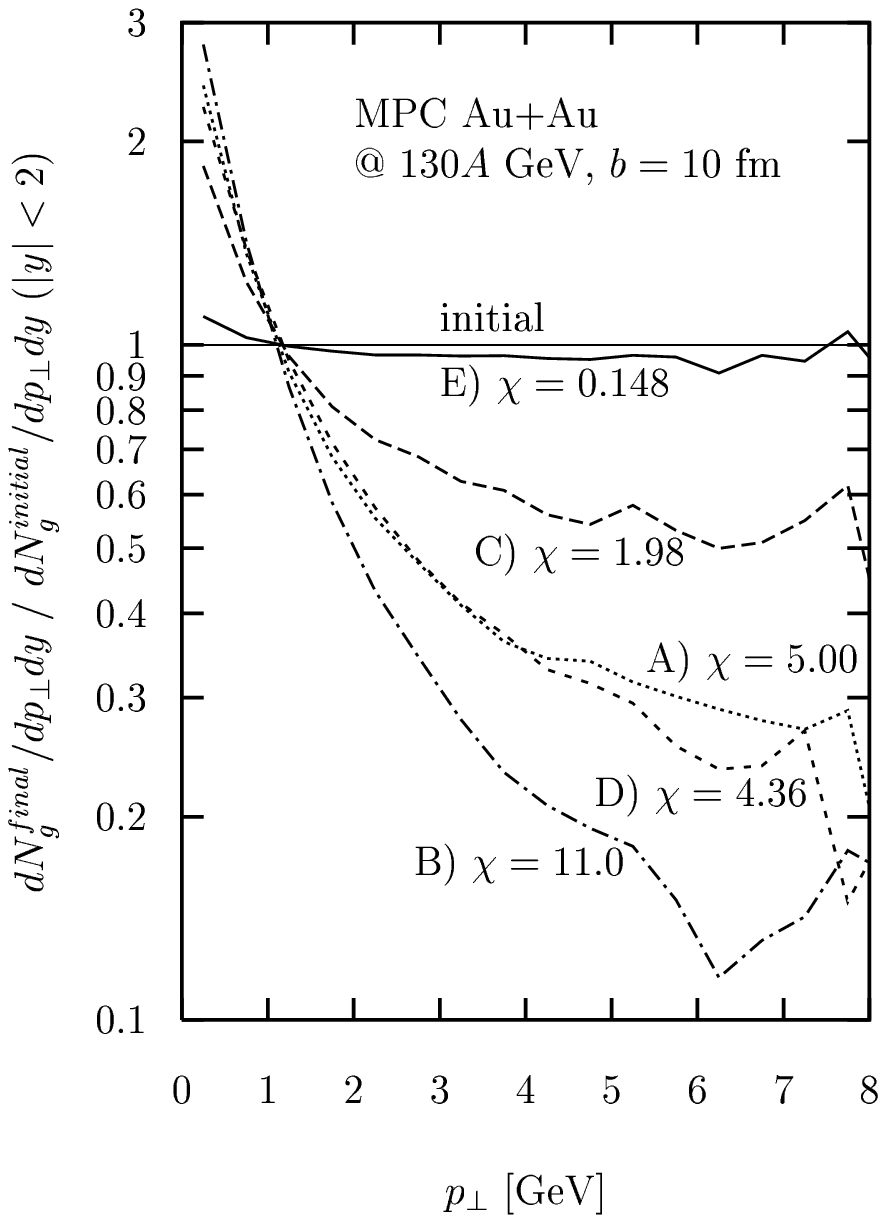}
    \epsfysize 6cm
    \epsfbox{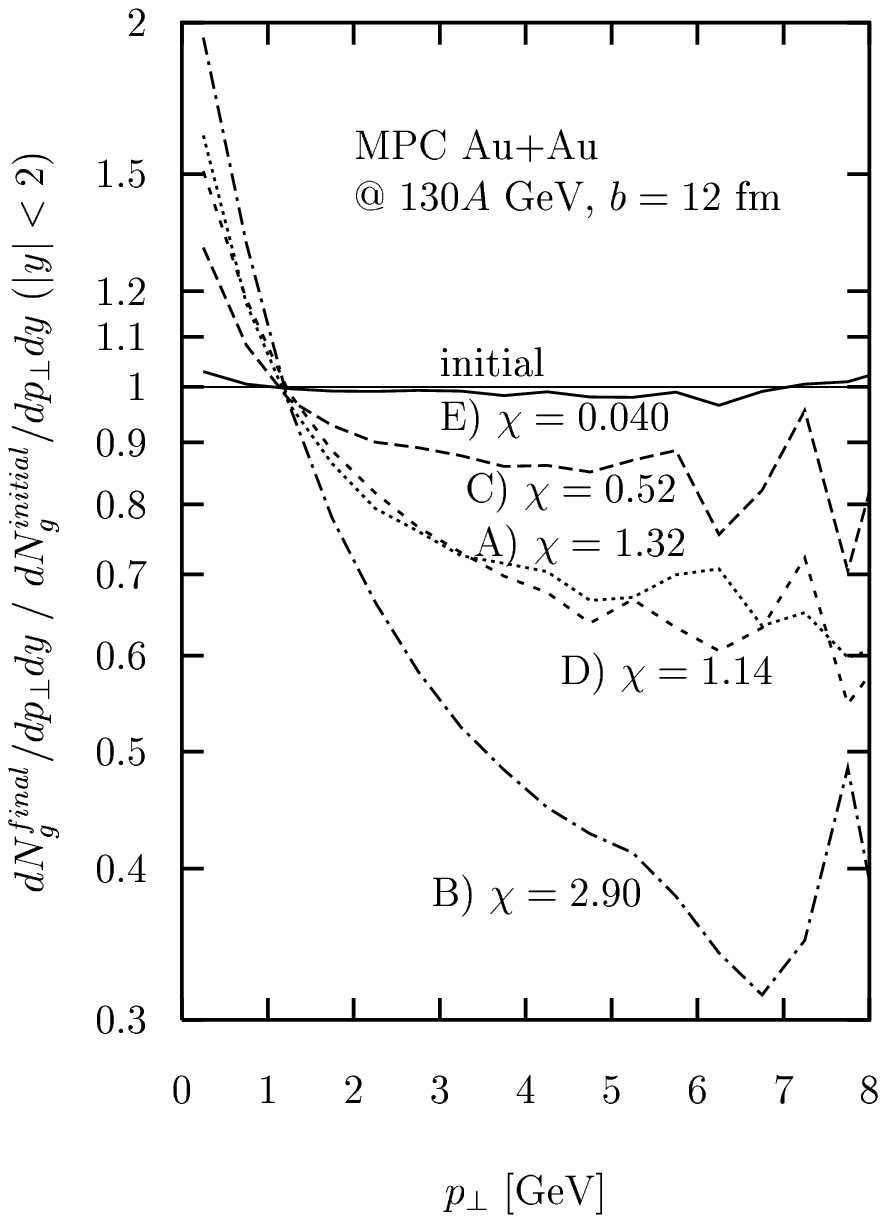}
}
\caption{
\footnotesize
Final gluon $p_\perp$ spectra relative to the thermal initial spectrum
are shown as a function of transport opacity
for Au+Au at $\sqrt{s} = 130A$ GeV
with $b=10$ fm (left) and $b=12$ fm (right).
Proportional to the charged hadron $p_\perp$ spectra
if the gluons are hadronized via local parton-hadron duality.
}
\label{Figure:10}
\label{Figure:pt_vs_sigma-2}
\end{figure}

\begin{figure}[hp]
\center
\leavevmode
\hbox{
    \epsfysize 8cm
    \epsfbox{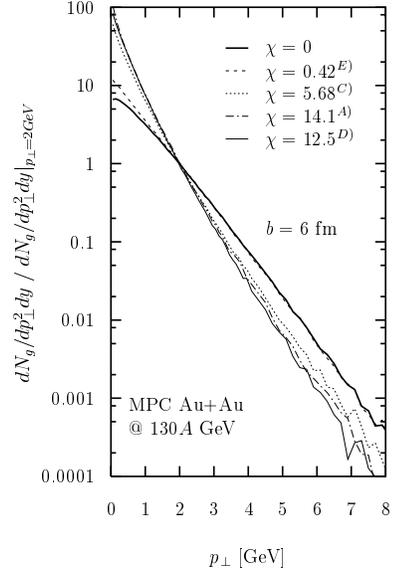}
}
\caption{
\footnotesize
Final gluon $p_\perp$ spectra as a function of transport opacity
are shown
for Au+Au at $\sqrt{s} = 130A$ GeV
with $b=6$ fm and
with all curves normalized to 1 at $p_\perp = 2$ GeV.
This shows the quenching at high $p_\perp$ relative to $p_\perp = 2$ GeV.
Also proportional to the charged hadron $p_\perp$ spectra
if the gluons are hadronized via local parton-hadron duality.
Preliminary STAR 
data\protect\cite{Dunlop:2001vh} suggest that
this ratio may reach $10^{-4}$ in the $p_\perp\sim 5-6$ GeV/c range.
}
\label{Figure:11}
\label{Figure:pt_vs_sigma_scaled}
\end{figure}

Figures \ref{Figure:pt_vs_sigma-1}-\ref{Figure:pt_vs_sigma_scaled}
show the final gluon $p_\perp$ spectra from MPC
as a function of transport opacity
and impact parameter. To show more clearly the degree of quenching
due to multiple elastic collisions,
we plot in Figs. \ref{Figure:pt_vs_sigma-1} and \ref{Figure:pt_vs_sigma-2}
the ratio of
the final spectra to the initial for $b=6,$ 8, 10, and 12 fm.
With HIJING initial densities and pQCD elastic cross sections,
$\chi \sim 0.02-0.2$ is too small to
produce more than $\sim 10$\% quenching. As
 we increase the transport opacity,
quenching 
of the $p_\perp>2$ GeV/c range increases
 and by $\chi\sim 14-16$ it 
reaches a factor of ten suppression at $p_\perp > 6$ GeV.

While the quenching
depends on $\chi$ only,
the absolute yield is proportional to the initial $dN/d\eta$.
Hence from the absolutely normalized measured spectrum 
at a given centrality
one could extract both $\sigma_t$ and $dN(0)/d\eta$.
As seen in Fig. \ref{Figure:pt_vs_sigma_scaled},
the quenching at high $p_\perp$ is complemented by an
enhancement at low $p_\perp$. For clarity, we normalized all curves at 
$p_\perp=2$ GeV. While the STAR data\cite{Dunlop:2001vh} are too 
preliminary to show yet, the slope appears to be much steeper
than the computed gluon slopes because hadronization further softens the spectra as discussed later.

\subsection{Transport opacity dependence}

In this Section we provide a qualitative
 explanation
for the remarkable
invariance of the results on the actual angular dependence
of the differential cross section (\ref{Eq:cross_section}).
The reason that this simplification occurs is that for the $\mu$ 
and gluon energy range considered in these plots,
the transport opacity is actually high enough
that little memory of the {\em initial}
 gluon momentum direction remains after multiple scattering. 
Thus, isotropic scattering and $\mu=T$ forward scattering both lead to essentially a random reorientation
of all the gluons involved. 

\begin{figure}[hp]
\center
\leavevmode
\hbox{
    \epsfysize 6cm
    \epsfbox{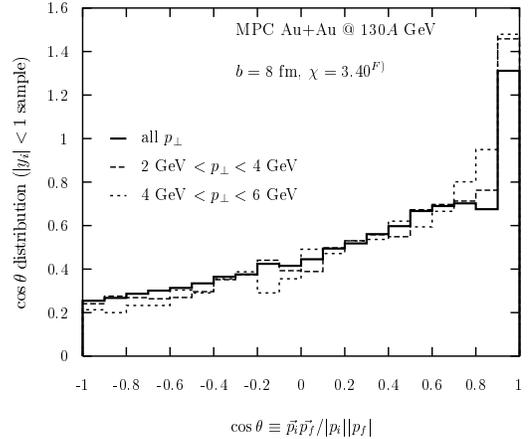}
}
\caption{
\footnotesize
Correlation between initial and final gluon momentum direction
is shown for Au+Au at $\sqrt{s}=130A$ GeV
with $b=8$ fm and transport opacity $\chi = 3.40^{F)}$.
}
\label{Figure:13}
\label{Figure:costheta}
\end{figure}

\begin{figure}[hp]
\center
\leavevmode
\hbox{
    \epsfysize 6cm
    \epsfbox{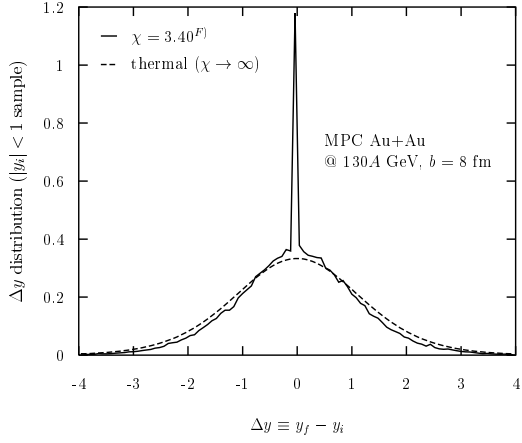}
}
\caption{
\footnotesize
Correlation between initial and final gluon rapidity is 
shown for Au+Au at $\sqrt{s}=130A$ GeV
with $b=8$ fm and transport opacity $\chi = 3.40^{F)}$.
}
\label{Figure:12}
\label{Figure:deltay}
\end{figure}

In Fig. \ref{Figure:costheta}, 
aside from a small $10\%$ delta function component due to the gluons
in the ``corona'' surface region that escape without rescattering,
the bulk of the minijets undergo enough 
rescatterings that their final direction is randomized. 
In Fig. \ref{Figure:deltay} we also show that
the rapidity shift of gluons in each transverse momentum interval considered
has the form close to one expected if local thermal equilibrium
occurred:
\bea
\frac{dP}{d\Delta y} &=&
\int dm_{\perp i}^2 dy_i 
dm_{\perp f}^2
m_{\perp i} \cosh y_i \;
m_{\perp f} \cosh (y_i + \Delta y)
\nonumber\\
&\times&
\frac{1}{4 T_i^3}\frac{1}{4 T_f^3}e^{-(m_{\perp i} \cosh y_i) / T_i}
e^{-[m_{\perp f} \cosh (y_i + \Delta y)] / T_f}
\nonumber\\
&=& \frac{\Delta y \cosh \Delta y - \sinh \Delta y}{\sinh^3 \Delta y} \ .
\eea

The randomization of momenta is however not sufficient to ensure
 the validity of local equilibrium necessary for the
applicability of nondissipative (Euler) hydrodynamics.
This  is proven by the dependence of the transverse
 momentum spectra and elliptic flow on the
{\em finite} opacity parameter itself.
In the hydrodynamic limit $\chi=\infty$ 
and the transport evolution is identical to the hydrodynamical evolution.
However, we
showed in detail in a previous study \cite{nonequil}
that the solutions of the transport equation still differ very much
from ideal  hydrodynamics, for physically extreme $\chi\sim 20$ opacities.
While no covariant 3+1D Navier-Stokes solutions are yet known,
our transport solutions demonstrate the effects of dissipation through
their dependence on $1/\chi$. 

The invariance of the transport solutions to the angular distributions
for a fixed $\chi$ indicate however that we are not extremely far
from the local thermal though dissipative limit.
In particular, our high opacity solutions are far
from the Eikonal (Knudsen) type dynamics as considered 
in \cite{Gyulassy:2000gk}. 

A rough criterion for the validity of the Eikonal approximation
is that the angle between the initial and final parton momenta
in the laboratory frame satisfies
$\Delta \theta \ll 1$,
say $\Delta \theta < 0.3$.
For an energetic parton that undergoes $N$ elastic collisions,
this angle can be approximated in an analogous way to random walk
as

\be
\langle\Delta \theta\rangle_N
\approx \sqrt{N \langle \Delta\theta^2\rangle_{N=1}}
\sim \frac{\sqrt{s}}{{2E}} \sqrt{N \langle\Delta\theta_{cm}^2\rangle}
\label{Eq:deflection_angle}
\ee
because the angles transform as
$\Delta\theta \approx \Delta \theta_{cm} \sqrt{s}/{2E}$.
In the small angle approximation, we can use Eq. (\ref{Eq:transport_xs})
to estimate
$$
\langle\Delta \theta_{cm}^2\rangle
\approx \frac{\sigma_t(s)}{\sigma_0}
\;\; .
\nonumber
$$
Hence, for elastic collisions off typical thermal partons
($s\approx 6ET$, $\mu = T$),
the Eikonal approximation is valid if $\sigma_t(s)/\sigma_0< E/(15 T N)$.
For $N\sim 10$,
this condition is satisfied 
for $E > 20 T$.
Note that the total cross section 
does not depend on the parton energy
and therefore the
number of collisions is approximately independent of energy.
In Fig. \ref{Figure:costheta} we see
that the Eikonal limit
is only approached slowly  as the parton energy increases.

We conclude from these results
that the pattern of ``jet quenching'',
as observed at RHIC via the suppression of 
moderate transverse  momentum particles
and the saturation of elliptic flow above some critical $p_\perp$,
can be reproduced if sufficiently high transport opacities 
are postulated\ with any angular distribution.

\section{Hadron Spectra}
\label{Section:results}
The results in the previous Section pertain only to partons.
To compare with experimental results, we must adopt a model of
 hadronization.
Here we compute the hadronic observables from
two different hadronization schemes.

\subsection{Hadronization via local parton-hadron duality}

The simplest hadronization scheme is 
based on the idea of local parton-hadron duality.
If as in
Ref. \cite{Eskola:2000fc}, we 
assume that each gluon gets converted
to a pion with equal probability for the three isospin states, 
then we may
approximate the transverse momentum distribution of
negative charged hadrons
roughly as
\be
f_{h^-}(\vec p_\perp) \approx f_{\pi^-}(\vec p_\perp) = \frac{1}{3} f_{g}(\vec p_\perp).
\label{Eq:hminus_from_glue}
\ee

\begin{figure}[hp]
\center
\leavevmode
    \epsfysize 6cm
    \epsfbox{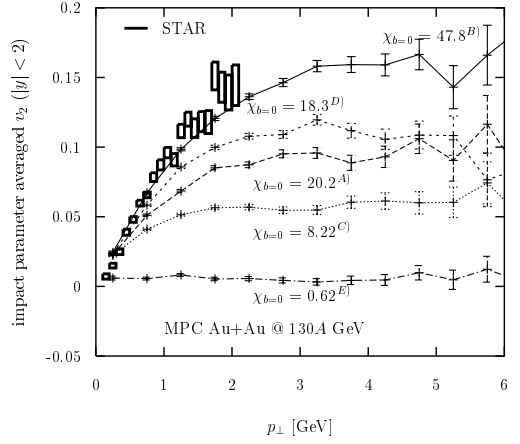}
\caption{
\footnotesize
Impact parameter averaged gluon elliptic flow is shown
as a function of $p_\perp$ for Au+Au at $\sqrt{s}=130A$ GeV
with transport opacities $\chi_{b=0} = 0.62^{E)}$, 8.22$^{C)}$,
18.3$^{D)}$,
20.2$^{A)}$, and 47.8$^{B)}$
for $b=0$.
Identical to the charged hadron elliptic
flow if the gluons are hadronized via local parton-hadron duality.
STAR 
data\protect\cite{STARv2} below 2 GeV/$c$ are shown.
Preliminary STAR data\protect\cite{Snellings} suggest that
$v_2\sim 0.15-0.17$ may saturate in the $2<p_T<4$ GeV/$c$ range.
}
\label{Figure:7}
\label{Figure:minbias_v2}
\end{figure}

With the above prescription, elliptic flow does {\em not}
change during hadronization,
i.e., Figs. \ref{Figure:v2_g40mb}--\ref{Figure:v2_vs_mu}
show the negative hadron flow as well.
Furthermore, the negative hadron $p_\perp$ spectra can be
obtained from Figs. \ref{Figure:pt_vs_sigma-1}--%
\ref{Figure:pt_vs_sigma_scaled}
via simply dividing by 3.
Consequently,
the scaling (\ref{Eq:v2_pt_scaling}) holds for the negative hadron flow and
spectra as well.

In Fig. \ref{Figure:minbias_v2},
the elliptic flow data by STAR are
reproduced with a transport opacity $\chi_{b=0}=47.8$.
For an initial gluon density $dN_g(0)/d\eta=1000$,
this corresponds to $\sigma_t \approx 14$ mb,
i.e., to a total cross section of $\sigma_0 \approx 45$ mb with $\mu = T_0$.
If we took, on the other hand, 
the pQCD $gg$ cross section of 3 mb with $\mu=T_0$,
this opacity would correspond to an initial gluon density of
$dN_g(0)/d\eta \sim 15000$ that is contradicted by the much smaller
observed $dN_{ch}/d\eta\approx 600$.

The $p_\perp$ spectra provide a much stronger constraint on
the  initial gluon density
as their absolute magnitude is proportional to it.
At high opacities ($\chi > \sim 10$),
the need for high particle subdivisions poses a severe computational problem,
therefore
we can reliably compute particle spectra for semicentral collisions only.
Nevertheless,
the data measured by STAR in central collisions,
where quenching due to parton energy loss is expected to be maximal,
provides an important lower bound on the particle yields.
Figure \ref{Figure:pt_vs_sigma_scaled} shows that
the elastic transport opacities $\chi < \sim 20$
considered here are compatible with this lower bound.

On the other hand, the cascade semicentral results are very much
above the preliminary STAR spectra\cite{Dunlop:2001vh} for central collisions.
The problem is that the fragmentation of quarks and gluons
generally soften considerably the high-$p_\perp$ spectra as we show in the next
Section.

\subsection{Hadronization via independent fragmentation}

The next simplest
 hadronization scheme is the fragmentation
 of gluons as independent
jets.
We
consider here only the $g\to \pi^{\pm}$ channel
with the next-to-leading-order
fragmentation function taken from
 Ref. \cite{Binnewies:1995ju}.
We took
the scale factor $s \equiv \log(Q^2)/\log(Q_0^2)$ to be zero
because  the initial HIJING gluon distribution is already
``self-quenched''
due to initial and final state radiation.
Also, 
since we do not consider the contribution of low-$p_\perp$
soft multiparticle production (beam jet fragmentation), 
we consider from now on only hadrons with $p_\perp > 2$ GeV.

\begin{figure}[hp]
\center
\leavevmode
    \epsfysize 6cm
    \epsfbox{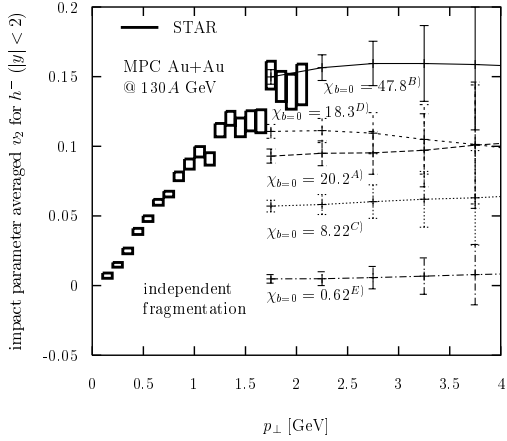}
\caption{
\footnotesize
Impact parameter averaged negative hadron elliptic flow
as a function of $p_\perp$\ is shown for Au+Au at $\sqrt{s}=130A$ GeV
with transport  opacities $\chi_{b=0} = 0.62^{E)}$, 8.22$^{C}$,
18.3$^{D)}$,
20.2$^{A)}$, and 47.8$^{B)}$
at $b=0$
and hadronization via independent fragmentation.
The $p_\perp < 2$ GeV region is not plotted because it is dominated
by soft contributions not addressable via pQCD jet fragmentation physics.
STAR 
data\protect\cite{STARv2} below 2 GeV/$c$ are shown.
Preliminary STAR data\protect\cite{Snellings} suggest that
$v_2\sim 0.15-0.17$ may saturate in the $2<p_T<4$ GeV/$c$ range.
}
\label{Figure:14}
\label{Figure:minbias_v2_frag}
\end{figure}

Figure \ref{Figure:minbias_v2_frag} shows the final impact-parameter-averaged
negative hadron flow as a function of the transport opacity.
The flow pattern and the magnitude of the flow are much the same as
for partons in Fig. \ref{Figure:minbias_v2}.
Hence, we get the same constraint on the initial parameters
as for hadronization via local parton-hadron duality.
In the $p_\perp < 2$ GeV region this simple calculation does not reproduce
the data because it does not include contributions coming from soft physics.

\begin{figure}[hp]
\center
\leavevmode
\hbox{
    \epsfysize 6cm
    \epsfbox{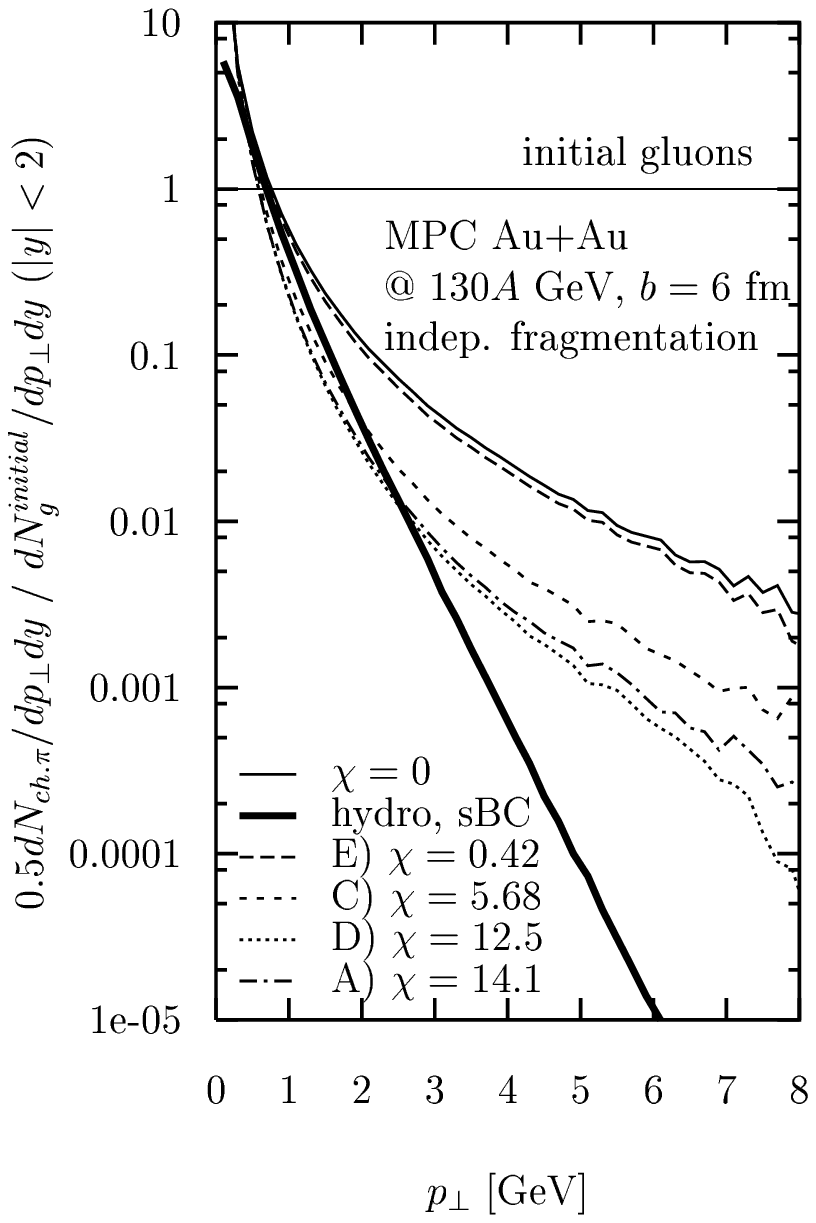}
    \epsfysize 6cm
    \epsfbox{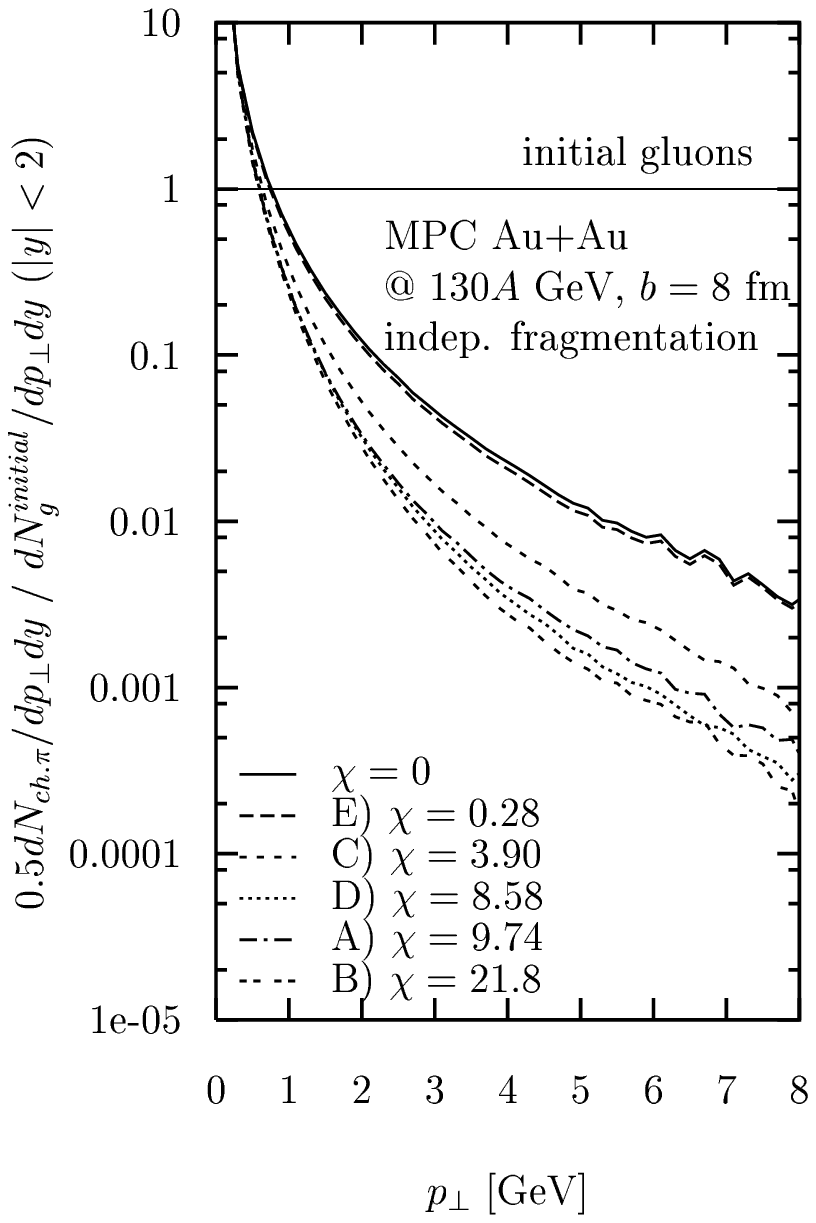}
}
\caption{
\footnotesize
Final negative hadron
$p_\perp$ spectra via independent fragmentation are shown
relative to the thermal initial gluon spectrum
for Au+Au at $\sqrt{s} = 130A$ GeV
with $b=6$ fm (left) and $b=8$ fm (right).
The ideal hydrodynamics result in the left figure is taken
from {\protect\cite{Kolbhydro}} with the so called sBC initial conditions.
It was extrapolated beyond $p_\perp = 3$ GeV using an exponential fit
to the $dN/p_\perp dp_\perp$ distribution between 2 and 3 GeV.
}
\label{Figure:15}
\label{Figure:pt_frag_vs_sigma-1}
\end{figure}

\begin{figure}[hp]
\center
\leavevmode
\hbox{
    \epsfysize 6cm
    \epsfbox{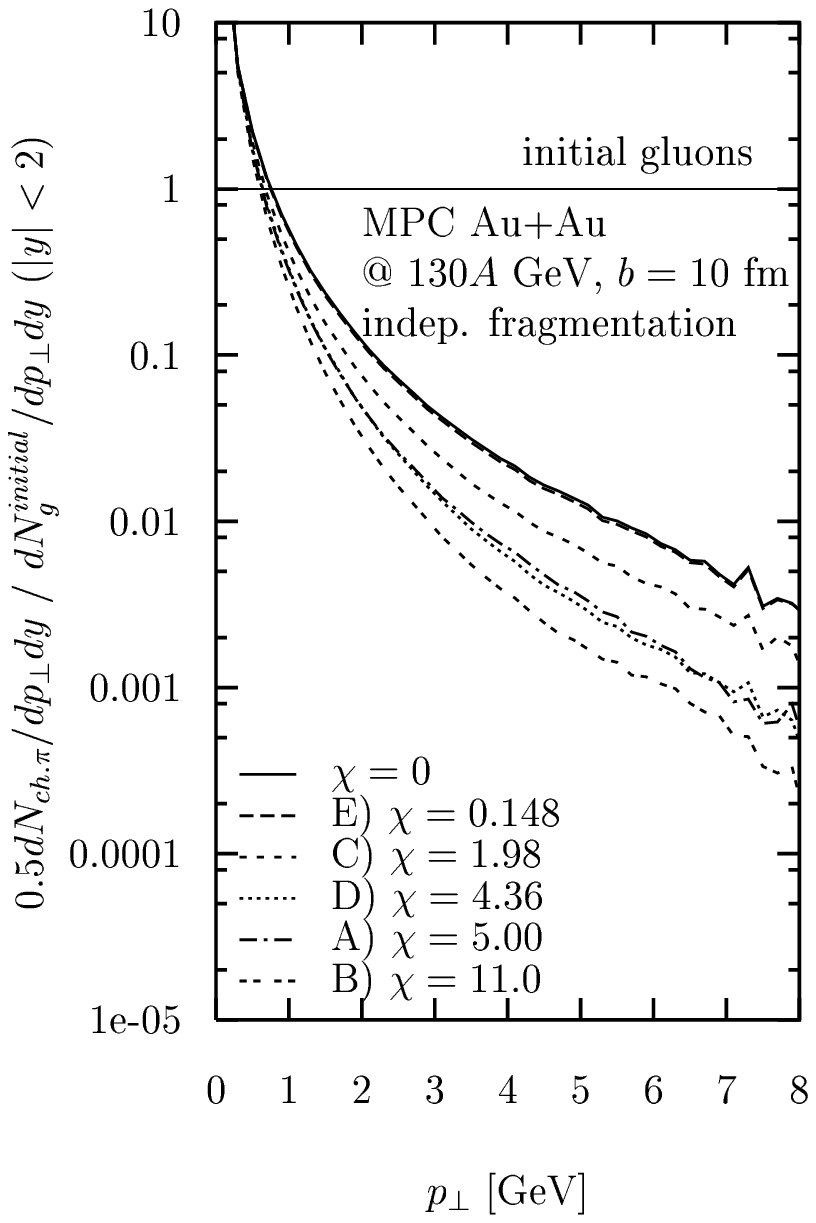}
    \epsfysize 6cm
    \epsfbox{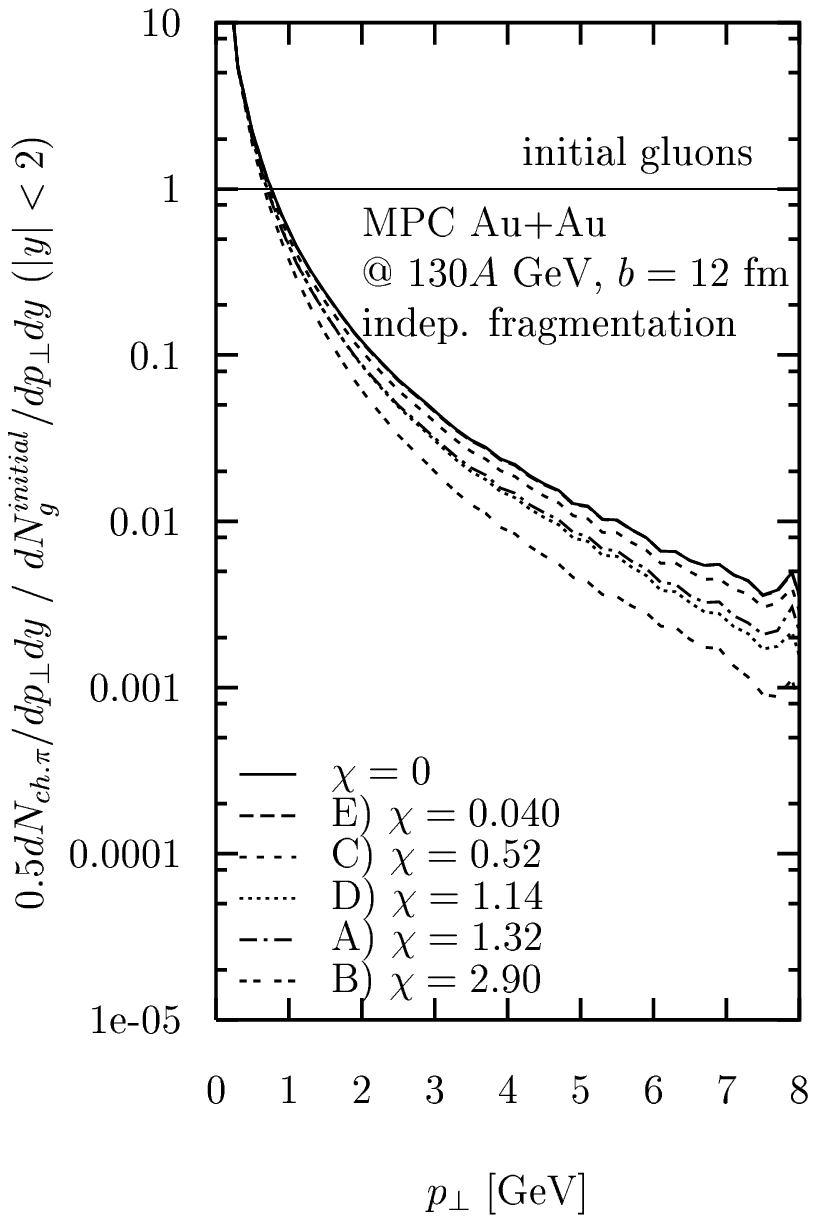}
}
\caption{
\footnotesize
Final negative hadron
$p_\perp$ spectra via independent fragmentation
are shown
relative to the thermal initial gluon spectrum
for Au+Au at $\sqrt{s} = 130A$ GeV
with $b=10$ fm (left) and $b=12$ fm (right).
}
\label{Figure:16}
\label{Figure:pt_frag_vs_sigma-2}
\end{figure}

\begin{figure}[hp]
\center
\leavevmode
\hbox{
    \epsfysize 8cm
    \epsfbox{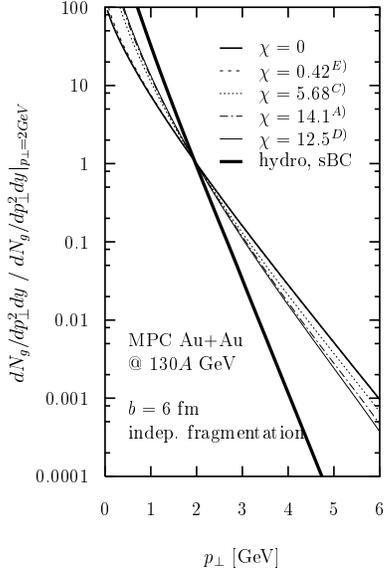}
}
\caption{
\footnotesize
Final negative hadron
$p_\perp$ spectra via independent fragmentation
are shown
relative to the thermal initial gluon spectrum
for Au+Au at $\sqrt{s} = 130A$ GeV
with $b=6$ fm and
with all curves normalized to 1 at $p_\perp = 2$ GeV.
This shows the quenching at high $p_\perp$ relative to $p_\perp = 2$ GeV.
The ideal hydrodynamics result is taken
from {\protect\cite{Kolbhydro}} with the so called sBC initial conditions,
and was extrapolated beyond $p_\perp = 3$ GeV using an exponential fit
to the $dN/p_\perp dp_\perp$ distribution between 2 and 3 GeV.
Preliminary STAR 
data\protect\cite{Dunlop:2001vh} suggest that
this ratio may reach $10^{-4}$ in the $p_\perp\sim 5-6$ GeV/c range.
}
\label{Figure:17}
\label{Figure:pt_vs_sigma_frag_scaled}
\end{figure}

The $p_\perp$ spectra of charged hadrons are shown 
as a function of transport opacity and impact parameter
in 
Figs. \ref{Figure:pt_frag_vs_sigma-1} and \ref{Figure:pt_frag_vs_sigma-2}.
In addition to quenching because of energy loss,
the final pion spectra are further quenched 
due to independent fragmentation.
With this additional quenching, the parton cascade
results approach the preliminary STAR data\cite{Dunlop:2001vh}, 
as indicated in Fig.
\ref{Figure:pt_vs_sigma_frag_scaled}, only for rather extreme
$\sigma_0(\mu=T) \approx 100$ mb if
HIJING $dN_g/dy=210$ is assumed, or if $\sigma_0 \approx 25$ mb and EKRT
$dN_g/dy=1000$ is assumed. These elastic cross sections exceed the conventional
few mb pQCD cross sections at this scale by at least an order of magnitude.

\section{Conclusions}

The MPC parton cascade technique was applied
to solve the covariant Boltzmann
transport numerically and compute new observables at RHIC.
Our focus was on the preliminary 
 differential elliptic flow and charged hadron moderate $p_\perp>2 $ GeV/c
spectra.
We compared results using  two different
hadronization schemes: 
independent fragmentation and  local parton-hadron
duality.

Our  main result is  that
if only elastic scattering is taken into account in the covariant Boltzmann equation, extremely large densities and/or elastic parton cross sections,
$\sigma_{tot} dN/d\eta \sim 80$ times the HIJING estimate,
are needed to reproduce the elliptic flow data\cite{Snellings}. 
Hadronization via local parton-hadron duality fails to reproduce the rapidly
falling  high-$p_\perp$ spectra.
However, independent fragmentation of our MPC solutions
compare well to
the charged hadron $p_\perp$ spectra
when rather  large 
elastic transport opacities are postulated.

The solutions clearly  demonstrate  how finite (even extreme) 
reaction rates
in $A+A$ lead to  major deviations from 
ideal hydrodynamic transverse 
flow effects at  transverse momentum $p_T>2$ GeV.
The pattern of quenching found with MPC is surprisingly  similar to 
that obtained in the two component
model of GVW\cite{Gyulassy:2000gk}.
The main difference between the high opacity MPC solutions reported 
here and the low opacity results of GVW is that
the latter include radiative energy loss in an Eikonal formalism
joined to a parametrized phenomenological ``hydrodynamic'' component.

It is known that radiative energy loss of ultrarelativistic partons
is much larger than elastic energy loss in a medium for a fixed
cross section.
In GVW a similar  quenching pattern was obtained 
with more modest
initial densities $ dN_g/dy\sim 500$ and  small pQCD elastic rates
because the induced gluon radiation associated with  multiple elastic
collisions
is large enough to compensate for the small elastic transport 
opacity in that case.
In MPC the same level of quenching is  achieved
only when the elastic opacity is increased
artificially by an order of magnitude.
Therefore, the present study confirms the 
expectation that elastic scattering alone is not enough to
generate the degree of collectivity observed now at RHIC.

\section{Outlook}
\label{Section:outlook}

The results presented here underscore the 
urgent need to develop practical convergent algorithms
to  incorporate inelastic $2\leftrightarrow 3$ processes.
Preliminary work in Ref. \cite{molnar:QM99} indicated
a rather slow convergence towards Lorentz covariance
using the particle subdivision technique. 
Unlike the $\ell^{-1/2}$ convergence of $2\to2$ transport solutions,
a much slower rate of convergence $\propto  \ell^{-1/5}$ 
is expected with the parton subdivision method used
to retain Lorentz covariance of  $2\leftrightarrow 3$ processes.

In addition, a more powerful covariant approximation 
to Boltzmann transport theory
may be needed
to overcome the overwhelming computational difficulty
in the high opacity regime for central collisions.
We found that even for the case of elastic scattering,
particle subdivisions up to $1000$ are required to maintain covariance
and stabilize the final spectra.

Finally, 
we note that all results in this paper pertain
to slowly varying, smooth initial conditions.
In Ref. \cite{turb},
it was suggested that copious minijet production
may  induce large (nonstatistical) local fluctuations
that could evolve in a turbulent manner.
A transport study of differential elliptic flow
and jet quenching in such inhomogeneous initial conditions
would be interesting  to compare to the 
presently known hydrodynamic and Boltzmann solutions.

\section{Acknowledgments}

We acknowledge the Parallel Distributed Systems Facility
at the National Energy Research Scientific Computing Center
for providing computing resources.
We also thank RMKI/KFKI for hospitality during the completion of this work.

This work was supported by the Director, Office of Energy Research,
Division of Nuclear Physics of the Office of High Energy and Nuclear Physics
of the U.S. Department of Energy under contract No. DE-FG-02-93ER-40764.
M. G. also was partially supported by Collegium Budapest. 

\appendix
\section*{Comparison of elliptic flow definitions}
\label{App}

The main reason that definition (\ref{Eq:v2MPC}) gives
a very similar elliptic flow to definition (\ref{Eq:v2STAR}),
contrary to the opposite observation from hydrodynamics,
is the difference between the hydro and the cascade $v_2(b)$ shapes.
Hydrodynamical models predict an increasing $v_2$ out to $b_{m}\approx 12-13$ fm,
while the cascade $v_2$ peaks at $b_{m}\approx 8$ fm.
The $b < b_{m}$ region where $v_2$ is small
gets a {\em larger} weight in the STAR definition,
while the $b > b_{m}$ region where $v_2$ is as well small
gets a {\em smaller} weight.
The former effect tends to reduce elliptic flow, while the latter
tends to increase it.
In the case of hydrodynamics,
the first effect dominates,
while for the cascade,
the second effect turns out to be larger.

We illustrate this with a simple analytic calculation.
The impact parameter dependence of elliptic flow can be fitted with the general
form
\be
v_2(b)
= K \left(\frac{b}{B}\right)^{c}
  \left(1-\frac{b}{B}\right)^{d} \ ,
\ee
while the particle spectrum is approximately linear in the $b=2-12$ fm region of interest
\be
\frac{dN}{dydp_\perp}(b) = C \left(1-\frac{b}{B'}\right) \ .
\ee
Here the parameters $K$, $B$, $B'$, $c$, $d$, and $C$ in general depend on $p_\perp$.

It is easy to show that for the above functions
definition (\ref{Eq:v2MPC}) gives an elliptic flow
$v_2^{av} = 2K\; \Gamma(c+2)\Gamma(d+1) / \Gamma(c+d+3)$,
while definition (\ref{Eq:v2STAR}) yields
$v_2^{STAR} = 6K\; \Gamma(c+2)\Gamma(d+2) / \Gamma(c+d+4)$,
provided we assume $B=B'$ and integrate up to $b_{max}=B$ in both cases.
Therefore,
\be
\frac{v_2^{STAR}}{v_2^{av}} = \frac{3d+3}{c+d+3} \ ,
\label{Eq:v2ratio}
\ee
which is larger than one, if and only if $2d >c$.
For our cascade results,
the fits to $v_2(b)$
give $d/c\sim 0.6-1.4$
[$c(p_\perp)\sim 1.1-3.3$, $d(p_\perp)\sim 1.0-4.5$],
therefore we have $v_2^{STAR}(p_\perp) > v_2^{av}(p_\perp)$.
On the other hand, hydrodynamics gives an approximately linear $v_2(b)$,
i.e., $c=1$ and $d=0$,
which results in $v_2^{STAR}(p_\perp) < v_2^{av}(p_\perp)$.

We found essentially
the same when taking an exponential fit to $dN/dp_\perp(b)$
instead of a linear one.

Equation (\ref{Eq:v2ratio}) yields $v_2^{STAR}/v_2^{av} \sim 1.1-1.5$ for the
cascade (depending on $p_\perp$ and initial conditions).
However, in reality $B'$ is {\em smaller}
than $B$ (typically $B'\approx 11-13$ fm, while $B\approx 13-16$ fm),
which influences $v_2^{STAR}$.
Furthermore, the upper limit of integration $b_{max}= 12$ fm
is also smaller than $B$.
This affects primarily $v_2^{av}$,
through the normalization constant $1/b_{max}^2$.
The integrals in both definitions are to a large degree insensitive to 
variations of $b_{max}$
because the integrands cut off naturally at large $b$.

To illustrate these effects, we repeat the previous analytic calculation with
$b_{max} \equiv x B$ for $v_2^{av}$,
while $B' \equiv y B$ and $b_{max} \equiv x' B'$ for $v_2^{STAR}$.
The integrals yield
\bea
&&v_2^{STAR}(x',y)= 6K \frac{y B_{yx'}(c+2, d+1)- B_{yx'}(c+3, d+1)}{y^3 x'^2(3-2x')}
\nonumber\\
&&v_2^{av}(x)= 2K\frac{B_x(c+2, d+1)}{x^2} \ ,
\label{Eq:v2ratio_b}
\eea
where $B_z(a,b) \equiv \int_0^z dt\; t^{a-1}(1-t)^{b-1}$ is the incomplete beta function.
Thus,
\bea
&&\frac{v_2^{STAR}(1-x',1-y)}{v_2^{STAR}(1,1)}
= 1+\frac{2d-c}{d+1}\; y+O(x'^2)+O(y^2)
\nonumber \\
&&\frac{v_2^{av}(1-x)}{v_2^{av}(1)}
=1+2x+O(x^{\min\left\{2,d+1\right\}}) \ ,
\eea
i.e., the leading correction to $v_2^{av}$ comes at first order in $(b_{max}-B)$,
while the leading correction to $v_2^{STAR}$ comes at first order
in $(B' - B)$.

For parameters and initial conditions considered in this study,
the corrected formulas (\ref{Eq:v2ratio_b})
yield $v_2^{STAR}/v_2^{av} \sim 0.9-1.05$
(with $x' = 1$).
The uncertainty of $v_2^{STAR}$ 
due to the unknown experimental cutoff $b_{max}$ (i.e., $x'$) is less than 5\%.
For the hydro ($c=1$, $d=0$), this analysis gives
a much smaller ratio
$v_2^{STAR}/v_2^{av} = 3 (1-x')(1- y) / 4(1-x) \approx 0.75$.


%
%

\newpage


\newpage

\newpage

\newpage

\newpage

\newpage

\newpage

\newpage

%
%

\newpage

\begin{table}
\begin{center}
\begin{tabular}{|c|c|c||c|c|c|c|}
\hline
\multicolumn{3}{|l||}{ {\bf A)} $\sigma_0 = 100$ mb,  $T_0/\mu = 1$} &
\multicolumn{4}{l|}{ {\bf B)} $\sigma_0 = 100$ mb,  $T_0/\mu = 0$} \\ 
\hline
$b$ [fm] & $\langle n \rangle$ & $\chi$ &
$b$ [fm] & $\langle n \rangle$ & \LCT{$\chi$} \\
\hline
\hline
0 & 33.0 & 20.2 &
0 & 35.8 & \LCT{47.8} \\
\hline
2 & 31.7 & 19.4 &
2 & 34.3 & \LCT{45.8} \\
\hline
4 & 28.1 & 17.2 &
4 & 30.2 & \LCT{40.2} \\
\hline
6 & 23.0 & 14.1 &
6 & 24.0 & \LCT{32.0} \\
\hline
8 & 15.9 & 9.74 &
8 & 16.3 & \LCT{21.8} \\
\hline
10 & 8.16 & 5.00 &
10 & 8.23 & \LCT{11.0} \\
\hline
12 & 2.15 & 1.32 &
12 & 2.18 & \LCT{2.90} \\
\hline
\hline
\hline
\multicolumn{3}{|l||}{ {\bf C)} $\sigma_0 = 40$ mb,  $T_0/\mu = 1$} &
\multicolumn{4}{l|}{ {\bf D)} $\sigma_0 = 40$ mb,  $T_0/\mu = 0$} \\ 
\hline
$b$ [fm] & $\langle n \rangle$ & $\chi$ &
$b$ [fm] & $\langle n \rangle$ & \LCT{$\chi$} \\
\hline
\hline
0 & 13.4 & 8.22 &
0 & 13.7 & \LCT{18.3} \\
\hline
2 & 12.9 & 7.90 &
2 & 13.2 & \LCT{17.6} \\
\hline
4 & 11.4 & 6.98 &
4 & 11.6 & \LCT{15.5} \\
\hline
6 & 9.26 & 5.68 &
6 & 9.38 & \LCT{12.5} \\
\hline
8 & 6.37 & 3.90 &
8 & 6.44 & \LCT{8.58} \\
\hline
10 & 3.23 & 1.98 &
10 & 3.27 & \LCT{4.36} \\
\hline
12 & 0.86 & 0.52 &
12 & 0.86 & \LCT{1.14} \\
\hline
\hline
\hline
\multicolumn{3}{|l||}{ {\bf E)} $\sigma_0 = 3$ mb,  $T_0/\mu = 1$} &
\multicolumn{4}{l|}{ {\bf F)} various, $b=8$ fm}\\ 
\hline
$b$ [fm] & $\langle n \rangle$ & $\chi$ &
$\sigma_0$ [fm] & $T_0/\mu$ & $\langle n \rangle$ & \LC{$\chi$} \\
\hline
\hline
0 & 1.00 & 0.62 &
60 & 1.54 & 9.51 & \LC{3.68} \\
\hline
2 & 0.96 & 0.58 &
16 & 0 & 2.55 & \LC{3.40}\\
\hline
4 & 0.85 & 0.52 &
100 & 1.40 & 15.9 & \LC{6.86}\\
\hline
6 & 0.69 & 0.42 &
100 & 2.21 & 15.7 & \LC{3.88} \\
\hline
8 & 0.47 & 0.28 &
100 & 4.43 & 15.5 & \LC{1.44}\\
\hline
10 & 0.24 & 0.148 &
 & & \LCT{}\\
\hline
12 & 0.064 & 0.040 &
 & & \LCT{}\\
\hline
\end{tabular}
\end{center}
\caption[1]{
\footnotesize
Parameters and transport opacity for each transport solution
computed via MPC for the present study.
}
\label{Table:chi}
\end{table}

\end{document}